\documentclass[reprint,aps,prb,twocolumn,superscriptaddress]{revtex4-2}

\usepackage{graphicx}
\usepackage{amsfonts}
\usepackage{amsmath}
\usepackage{amssymb}
\usepackage{bbm}
\usepackage{blindtext}  
\usepackage{bm}
\usepackage{braket}
\usepackage{bbold}
\usepackage{color}
\usepackage{comment}
\usepackage{hyperref}
\usepackage{latexsym}
\usepackage{multirow}
\usepackage{subfigure}
\usepackage{soul}
\usepackage{tikz}
\usepackage{tabularx}
\usepackage{upgreek}
\usepackage{verbatim}
\usepackage{xcolor}
\usepackage{physics}
\usepackage{soul}
\usepackage{placeins}

\hypersetup{
    colorlinks=true,
    linkcolor=blue,
    urlcolor=blue,
    filecolor=blue,
    citecolor=blue
    }
\usepackage{siunitx}

\begin{document}

\title{Neural-quantum-state based downfolding of the three-band Emery model \\ for cuprates and nickelates}

\author{Hannah Lange}
\affiliation{Department of Physics and Arnold Sommerfeld Center for Theoretical Physics (ASC), Ludwig-Maximilians-Universit\"at M\"unchen, Theresienstr. 37, M\"unchen D-80333, Germany}
\affiliation{Munich Center for Quantum Science and Technology (MCQST), Schellingstr. 4, M\"unchen D-80799, Germany}

\author{Julius F. A. Tirpitz}
\affiliation{Department of Physics and Arnold Sommerfeld Center for Theoretical Physics (ASC), Ludwig-Maximilians-Universit\"at M\"unchen, Theresienstr. 37, M\"unchen D-80333, Germany}
\affiliation{Munich Center for Quantum Science and Technology (MCQST), Schellingstr. 4, M\"unchen D-80799, Germany}

\author{Annabelle Bohrdt}
\affiliation{Department of Physics and Arnold Sommerfeld Center for Theoretical Physics (ASC), Ludwig-Maximilians-Universit\"at M\"unchen, Theresienstr. 37, M\"unchen D-80333, Germany}
\affiliation{Munich Center for Quantum Science and Technology (MCQST), Schellingstr. 4, M\"unchen D-80799, Germany}

\date{\today}
\begin{abstract}
Understanding the physics underlying high-temperature superconductivity in cuprates and, more recently, infinite-layer nickelates has remained a central challenge in condensed-matter physics. We establish neural quantum states (NQS), specifically Hidden Fermion Determinant States (HFDS), as a scalable variational approach to the three-band Emery model of the copper- and nickel-oxide layers in these materials. After benchmarking HFDS against matrix product states on width-two geometries, we study ground states of fully two-dimensional (2D) systems of up to $10\times10$ unit cells ($300$ sites). We characterize the momentum-space distribution of dopants and find a pronounced electron-hole dichotomy similar to cuprate experiments. We further downfold the three-band model to effective single-band descriptions by constructing interacting Wannier functions. We consider a wide range of parameters -- from the charge-transfer regime relevant to cuprates to the Hubbard-Mott regime of nickelates, as well as systematic scans of the charge-transfer gap that has been demonstrated to impact the critical superconducting temperatures. Across all regimes, the effective model significantly deviates from the usual Fermi-Hubbard model: The typical ratio $U/t$ is enhanced, some parameters experience a significant doping dependence, and sizable terms beyond the conventional Hubbard model are present, most notably a density-assisted hopping $t_n$~\cite{Jiang2023}. Notably, in all effective models, $t_n$ has the largest contribution to particle-hole asymmetry, rather than next-nearest-neighbor hopping contributions. The effective parameters sensitively depend on the charge-transfer energy, doping, and interaction ratios. Our results establish HFDS as an efficient tool for studying the 2D Emery model and demonstrate that single-band descriptions can require interaction terms generated by the underlying multi-band models.
\end{abstract}

\maketitle
Since the discovery of cuprate-based high-temperature superconductivity more than four decades ago~\cite{Bednorz1986}, understanding the microscopic mechanisms responsible for superconductivity in these materials has remained one of the central unresolved problems in condensed matter physics. The single-band Fermi-Hubbard model is arguably the most extensively studied minimal model in this context~\cite{Arovas_2022}. However, a more microscopic description explicitly accounts for the relevant electronic degrees of freedom of these planes, namely the copper $3d$ orbitals and the oxygen $2p$ orbitals. This leads to the three-band Emery model on a Lieb lattice geometry, which contains one copper and two oxygen orbitals per unit cell~\cite{Emery1987}.

The widespread use of effective single-band models is largely motivated by the seminal work of Zhang and Rice~\cite{Zhang1988} which demonstrated that, under suitable conditions, the low-energy degrees of freedom of the three-band model can be described in terms of copper-centered Wannier orbitals (Zhang-Rice singlets). Combined with perturbative arguments, this construction leads to an effective single-band $t$-$J$ model, which itself corresponds to the strong-coupling limit of the Fermi-Hubbard model. These arguments were subsequently refined and extended in several works, including Refs.~\cite{Schuettler1992,Jefferson1992,Belinicher1993,Belinicher1994,Feiner1996}. 

More recently, an increasing body of theoretical and experimental evidence suggests that such single-band descriptions may not be sufficient to capture all relevant aspects of the physics of cuprate compounds~\cite{Jiang2023,Xu2024,Padma2025,Scheie2025,Li2025,lange2025hamiltonianreconstruction}. In particular, the reduction from a multi-band to an effective single-band model may generate additional interactions beyond the standard nearest-neighbor Hubbard Hamiltonian. Prominently, with a numerical construction of interacting Wannier functions based on matrix product state (MPS) calculations~\cite{Jiang2023}, density-assisted tunneling processes have regained attention~\cite{Marsiglio1990,Hirsch1991,Hirsch1995}. Determining the importance of such additional terms is therefore essential for assessing to what extent the conventional single-band Fermi-Hubbard model provides an adequate description of high-T$_c$ superconductors.

A complementary, materials-oriented perspective emphasizes how the critical temperatures $T_c$ vary with the charge-transfer energy across cuprate families~\cite{ocallaghan2026chargetransfergapsizeoxygen,Weber_2012,Ruan2016Relationshipbetweentheparentchargetransfergap,Wang2023Correlatingthecharge-transfergaptothemaximumtransitiontemperature,Yee2014charge-transferenergyinhole-dopedcuprates} or, likewise, the charge distribution in the CuO$_2$ planes between the $p$ and $d$ orbitals, $n^p/n^d$~\cite{Jurkutat2014,Rybicki2016}: cuprates with the highest critical temperatures tend to have a relatively small charge-transfer gap and high $n^p/n^d$~\cite{ocallaghan2026chargetransfergapsizeoxygen,Weber_2012,Ruan2016Relationshipbetweentheparentchargetransfergap,Wang2023Correlatingthecharge-transfergaptothemaximumtransitiontemperature,Yee2014charge-transferenergyinhole-dopedcuprates,Jurkutat2014,Rybicki2016}. Relating this Cu-O charge redistribution (and, more broadly, the charge-transfer energy $\Delta_{pd}$) to the parameters of effective single-band Hamiltonians is therefore an important step toward understanding which microscopic ingredients control $T_c$.

\begin{figure*}[t]
    \centering
    \includegraphics[width=0.95\textwidth]{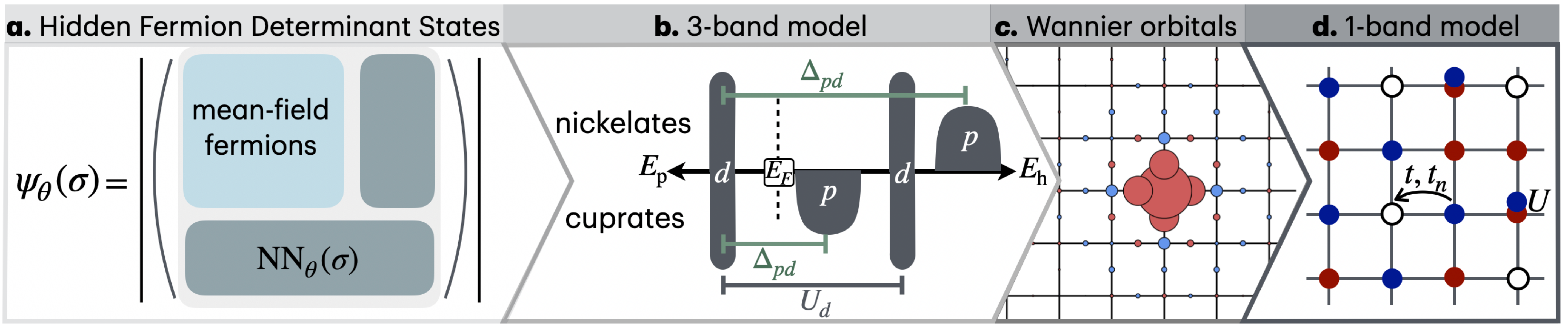}
    \caption{We employ Hidden Fermion Determinant States (HFDS, \textbf{a}) to model the three-band Emery model relevant to cuprate and nickelate materials~(\textbf{b}). This allows us to determine the interacting Wannier functions~(\textbf{c}), here shown for $\Delta_{pd}/t_{pd}=3.0$ and $\delta=1/8$, and to downfold to a single-band Fermi-Hubbard model~(\textbf{d}). 
    }
    \label{fig:1}
\end{figure*}

To address this question, it is useful to revisit the multi-band description from which the effective single-band models emerge. At the same time, substantial methodological advances have made increasingly accurate numerical investigations possible. Consequently, the three-band Emery model has experienced renewed interest in recent years: Tensor-network approaches have provided important insights into the ground-state properties and competing orders of the Emery model upon doping~\cite{White2015,Ponsioen2023,Jiang2023,zhang2026robustfluctuatingintertwinedcharge}. Quantum Monte Carlo methods have enabled studies of finite-temperature and ground-state properties~\cite{Huang2017,Chiciak2020,mei2025magneticelectronholeasymmetrycuprates,zhang2026robustfluctuatingintertwinedcharge}. Further perspectives have been obtained using density-matrix embedding methods and other embedding approaches~\cite{mei2025magneticelectronholeasymmetrycuprates,Kent2008}, as well as dynamical mean-field and cluster dynamical mean-field techniques~\cite{Tseng2025,ocallaghan2026chargetransfergapsizeoxygen}, with insights on the impact of the Emery model parameters on the superconducting transition temperature~\cite{Weber_2012,Ruan2016Relationshipbetweentheparentchargetransfergap,Wang2023Correlatingthecharge-transfergaptothemaximumtransitiontemperature,Yee2014charge-transferenergyinhole-dopedcuprates,ocallaghan2026chargetransfergapsizeoxygen}. Moreover, recent proposals demonstrate that the Emery model can be realized experimentally in cold atom setups~\cite{mccabe2026realizingmultiorbitalemerymodels,lange2026Emerycoldatoms}.

Despite this considerable progress, the larger local Hilbert space compared to its single-band counterpart, the presence of multiple orbitals and energy scales, and, for many numerical approaches, severe fermionic sign problems substantially increase the computational complexity compared to the single-band case. Developing scalable and controlled approaches to investigate the ground state properties of the Emery model is hence crucial not only for clarifying the microscopic physics of cuprates, but also for determining which effective interactions must be retained when constructing reduced single-band descriptions. Besides this, the Emery model is relevant to other unconventional superconductors, namely infinite-layer nickelates~\cite{Wang2024ExperimentalProgressinSuperconductingNickelates,Botana2020SimilaritiesandDifferences,Lin2021StrongSuperexchange,Chen2022Electronicstructureofsuperconductingnickelates} as well as other phenomena like altermagnetism~\cite{kaushal2024altermagnetismmodifiedlieblattice}. \\

In this work, we establish neural quantum states (NQS)~\cite{Medvidovic2024,lange2024reviewnqs,Carleo2017}, and in particular Hidden Fermion Determinant states (HFDS)~\cite{Moreno2022}, as an accurate and efficient approach for simulating the three-band Emery model (Fig.~\ref{fig:1}a). To our knowledge, this is the first time that NQS are employed for a three-band study. We first benchmark the accuracy of our approach against MPS on width-two geometries, where MPS provide highly accurate reference results. Then, we exploit the favorable scaling of our NQS approach for genuinely two-dimensional geometries to study square systems containing up to $N_\mathrm{cell}=L_x\times L_y=10\times10$ unit cells, corresponding to $300$ lattice sites, with periodic boundary conditions. 

The computational efficiency of NQS enables us to systematically explore an extended region of the Emery-model parameter space. In particular, we investigate regimes relevant to both hole- and electron-doped cuprates and nickelates, see Fig.~\ref{fig:1}b, allowing us to characterize the evolution of the ground-state properties across a broad range of physically relevant parameters and compare them between the two compounds. Furthermore, we systematically scan different values of the charge-transfer gap $\Delta_{pd}$ -- a quantity that has been shown to affect the critical temperatures for superconductivity~\cite{ocallaghan2026chargetransfergapsizeoxygen,Weber_2012,Ruan2016Relationshipbetweentheparentchargetransfergap,Wang2023Correlatingthecharge-transfergaptothemaximumtransitiontemperature,Yee2014charge-transferenergyinhole-dopedcuprates}. Finally, we construct interacting Wannier orbitals directly from the correlated ground states (Fig.~\ref{fig:1}c). By projecting the interacting multi-band problem onto the resulting low-energy degrees of freedom, we derive the corresponding effective single-band Hamiltonians, see Fig.~\ref{fig:1}d. This procedure follows Ref.~\cite{Jiang2023}, but we extend the analysis to systematic Emery parameter and doping scans. This procedure enables us to systematically identify and quantify the effective processes generated by the underlying multi-band physics: The effective $U/t$ is enhanced compared to the usual Fermi-Hubbard model; additional terms are present that generate the particle-hole asymmetry expected for the materials while next-neighbor hopping $t^\prime/t$ remains small; and the effective parameters are doping-dependent.

\section{Results}

The model studied in this work is the three-band Emery model, defined by
\begin{equation}
    \begin{aligned}
\hat{\mathcal{H}}_\mathrm{3b} = &  
-\frac{1}{2}\sum_{\nu,\nu^\prime} \sum_{\mathbf{i}\mathbf{j} , \sigma} t_{\nu\nu^\prime}^{\mathbf{ij}} (\hat{C}^{\dagger}_{\nu\mathbf{i}\sigma} \hat{C}_{\nu\mathbf{j}\sigma} + \text{h.c.}) \\
&+ \Delta_{pd} \sum_{\mathbf{i},\sigma} \hat{n}^p_{\mathbf{i}\sigma}
+ \sum_{\nu } U_\nu \sum_\mathbf{i} \hat{n}^\nu_{\mathbf{i}\uparrow} \hat{n}^\nu_{\mathbf{i}\downarrow}. \label{eq:3bandEmery} 
\end{aligned}
\end{equation}
with $\hat{C}^{(\dagger)}_{\nu\mathbf{i}\sigma}$ hole annihilation (creation) operators and $\nu=(d, p_x, p_y)$. The hole densities are given by \( \hat{n}^\nu_{\mathbf{i}\sigma} = \hat{C}^\dagger_{\nu\mathbf{i}\sigma} \hat{C}_{\nu\mathbf{i}\sigma} \). We define the copper-copper distance as the unit of length and consider $t_{pd}=:t_{pd}^\mathbf{ij}\vert_{\vert \mathbf{i}-\mathbf{j}\vert =0.5}$ for nearest-neighbor $d$-$p$ hopping; $t_{pp}^{(\prime)}:=t_{pp}^\mathbf{ij}\vert_{\vert \mathbf{i}-\mathbf{j}\vert =1/\sqrt{2}\,(1)}$ for (next-) nearest-neighbor $p$-$p$ hopping; and $t_{\nu\nu^\prime}^\mathbf{ij}=0$ otherwise. The charge-transfer energy is denoted by \( \Delta_{pd} \), while \( U_{d(p)} \) are the on-site Hubbard interactions for $d$ ($p$) sites. Eq.~\eqref{eq:3bandEmery} is defined in an orbital gauge where all considered hopping contributions are negative. Note that this shifts the quasi-momentum by $\mathbf{Q}=(\pi,\pi)$.  

At half-filling, there is one hole per unit cell, $N_h=L_xL_y$, predominantly residing at the copper sites due to $\Delta_{pd}$. Hole-doping corresponds to the addition of holes, $\delta=(N_h-L_xL_y)/(L_xL_y)>0$, electron doping to the removal $\delta<0$. 

Determining the appropriate parameters of the Emery model remains an active
area of research~\cite{vucicevic2026importanceeffectivecoulombinteractions,jacob2026conventionalemerymodelcrucial}. Estimates and parameterizations relevant to cuprate and nickelate compounds span a substantial range, with $t_{pp}/t_{pd}=0,\dots,0.5$; $U_d/t_{pd}=4.0,\dots,9.0$; $U_p/t_{pd}=0,\dots,6.0$ and  $\Delta_{pd}/U_{d}=0.3,\dots,0.5$ (cuprates) or $\Delta_{pd}/U_d=1.0,\dots,1.5$ (nickelates)~\cite{Hybertsen1989,Martin1996,Hanke2010,Cui2020,McMahan1988,White2015,Kent2008,MDopf1990,Jiang2023,Kowalski2021Oxygenholecontent,Tseng2025,Jiang2020criticalNatureoftheNiSpinState,Botana2020SimilaritiesandDifferences,Lin2021StrongSuperexchange,Chen2022Electronicstructureofsuperconductingnickelates,Wang2023Correlatingthecharge-transfergaptothemaximumtransitiontemperature}. Here, we will set $t_{pp}/t_{pd}=0.5$ and $t_{pp}^\prime=0.0$ if not stated otherwise.\\

The HFDS used in this work to simulate the Emery model augment the physical Hilbert space given by $N_v$ physical fermionic degrees of freedom with $N_h$ auxiliary, or hidden, fermionic degrees of freedom, followed by a projection onto the physical subspace~\cite{Moreno2022}. For a Fock-space configuration $\mathbf{\sigma}$, the resulting amplitudes can be written as
\begin{align}
\langle\mathbf{\sigma}|\psi_{\rm HFDS}\rangle
=
\det
\begin{pmatrix}
\mathbf{\sigma}\star\mathbf{U}_v
&
\mathbf{\sigma}\star\mathbf{U}_{vh,1}
\\
\mathbf{U}^{\mathbf{\theta}}_{vh,2}(\mathbf{\sigma})
&
\mathbf{U}^{\mathbf{\theta}}_h(\mathbf{\sigma})
\end{pmatrix},
\label{eq:HFDS}
\end{align}
where the lower block is parametrized by a neural network (here a convolutional neural network (CNN)) with trainable parameters $\mathbf{\theta}$~\cite{Chen2025Thesis,chen2025neuralnetworkaugmentedpfaffianwavefunctions}. Here, $\mathbf{\sigma}\star\mathbf{U}$ denotes the selection of the occupied rows of $\mathbf{U}$ according to the occupations in $\sigma$. We note that the evaluation of the determinant scales as $\mathcal{O}((N_v+N_h)^3)$. 

This approach and related ans\"atze have been successfully applied to single-layer $t$-$J$~\cite{lange2024quasiparticle,langeboehler2025simulating} and Fermi-Hubbard models~\cite{Moreno2022,Liu2024unifying,Luo2019,sharma2025comparingsymmetrizeddeterminantneural,chen2025neuralnetworkaugmentedpfaffianwavefunctions,roth2025superconductivitytwodimensionalhubbardmodel,gu2025solvinghubbardmodelneural}, including very recent work on lattices of up to $24\times24$ sites~\cite{rende2026superconductivitytthubbardmodel}. Furthermore, they have been established as powerful tools for bilayer fermionic models~\cite{lange2026mixDPfaffian}. Here, we demonstrate their efficiency and accuracy for the three-band model. Despite the apparent increase in complexity, the number of simulated fermions (which correspond to holes in our formulation) is not increased by a factor of three; instead, we have $N_v\approx L_xL_y$. Consequently, the dimension of the determinant remains of the same order as in single-band models, rendering its evaluation computationally comparable to that of the corresponding single-band ans\"atze.\\

In the following, we exploit this efficiency and present the results obtained with HFDS for the Emery model on lattices of up to $10\times10$ unit cells. We begin by benchmarking our energies against MPS results for width-$2$ systems. We then turn to square systems and investigate ground-state properties as a function of doping. Finally, we systematically study the downfolding to effective single-band models as a function of a range of Emery-model parameters relevant to cuprates and nickelates. The doping levels quoted in the text refer to the nominal doping, with the actual number of doped holes chosen as the nearest even integer.

\begin{figure}[t]
    \centering
    \includegraphics[width=0.49\textwidth]{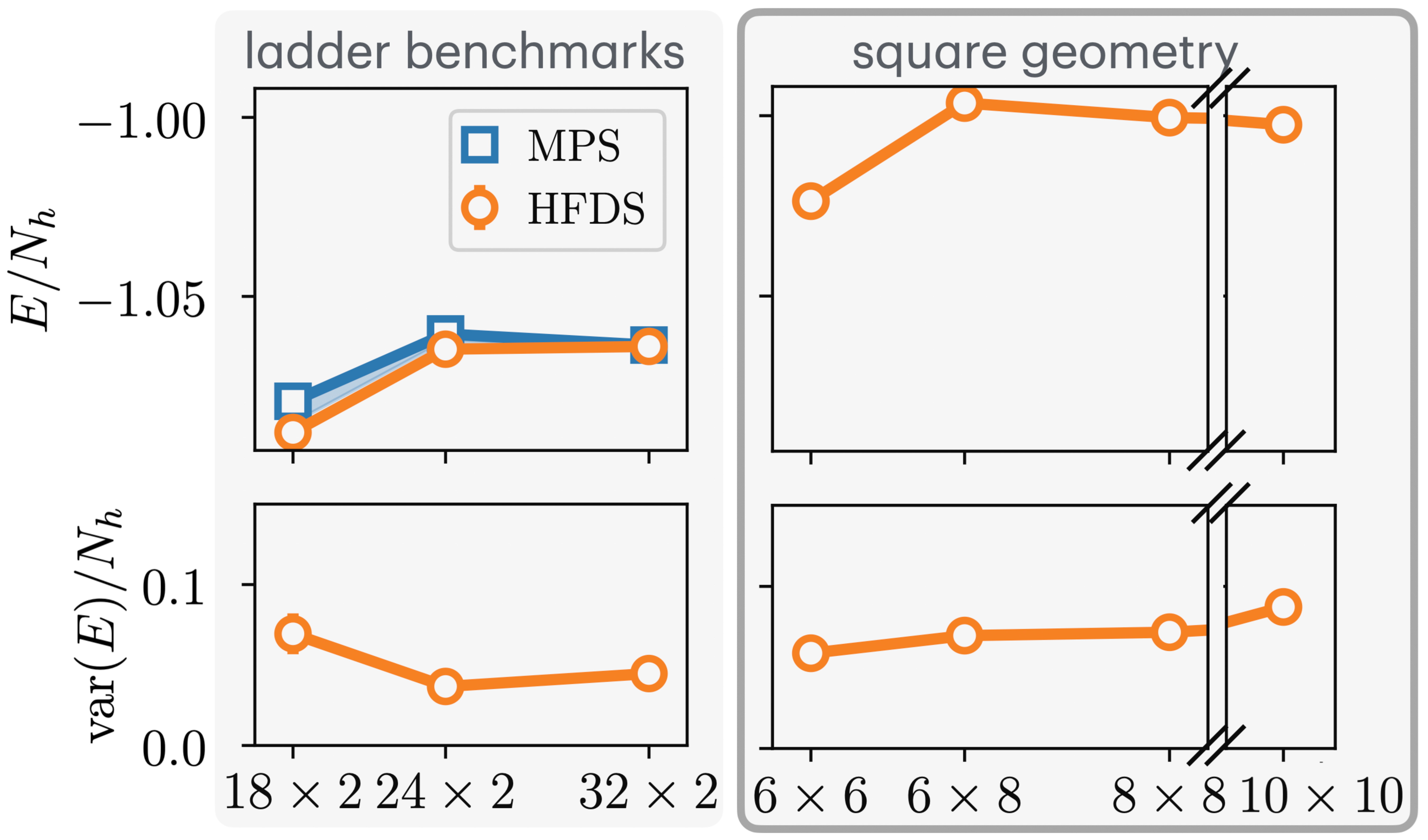}
    \caption{Energy $E$ and energy variance $\mathrm{var}(E)$ in units of $t_{pd}$ and per hole $N_h$ for $\delta=1/8$, $\Delta_{pd}/t_{pd}=3.5$ and $U_d/t_{pd}=6.0$. Left: We compare the HFDS (orange) and MPS (blue) energies  for long $L_x\times 2$ systems MPS calculations are feasible with periodic (open) boundaries in the long direction for HFDS (MPS). The shaded blue region indicates the estimate on the energy difference from the different boundary conditions. Right: HFDS results for square geometries with fully periodic boundaries.
    }
    \label{fig:benchmarks}
\end{figure}

\subsubsection{Energy comparison}
To assess the performance of the NQS ansatz, we benchmark its variational energies against MPS. We first consider narrow geometries of width $L_y=2$ for which MPS provide very accurate reference states. The resulting energies are shown in the left panel of Fig.~\ref{fig:benchmarks}. The MPS calculations employ a bond dimensions up to $\chi=4096$, corresponding to up to $4\chi^2\cdot 3L_xL_y=10^{9}$ parameters for the largest system with $32\times2$ unit cells; the NQS use $\sim 10^5$ parameters for the same system size, see Sec.~\ref{sec:methods}. We would like to point out that a direct comparison requires care because the MPS calculations are restricted to open boundary conditions (OBC) along the long direction, whereas the NQS calculations use periodic boundary conditions (PBC), consistent with the translational symmetry built into the CNN. This difference leads to visibly lower NQS energies (orange) compared to the MPS results (blue) for the smaller systems. For the larger systems, the boundary-condition effect becomes negligible. We estimate the contribution of the two missing bonds in the $x$ direction from the change in the MPS energy when changing $L_x$ to $L_x\pm 1$, which defines the blue-shaded energy range. After accounting for this boundary correction, the NQS and MPS energies are in good agreement across all system sizes. Moreover, the NQS energy variance remains approximately constant, with $\mathrm{var}(E)/N_h<0.1$, indicating stable convergence of the variational optimization. \\

The main advantage of the NQS is that it is not restricted to narrow geometries and can therefore be applied directly to square systems, as shown in the right panel of Fig.~\ref{fig:benchmarks}. The energy variance per hole for these systems is comparable to that obtained for the narrow geometries, further supporting the quality of the NQS representations also in the fully two-dimensional regime. Overall, these results demonstrate that our NQS provide competitive energies (with a substantially smaller number of variational parameters) while extending naturally to geometries beyond the quasi-one-dimensional regime. This allows us to study physically relevant observables next.

\subsubsection{Momentum-dependent densities}
\begin{figure}[t]
    \centering
    \includegraphics[width=0.49\textwidth]{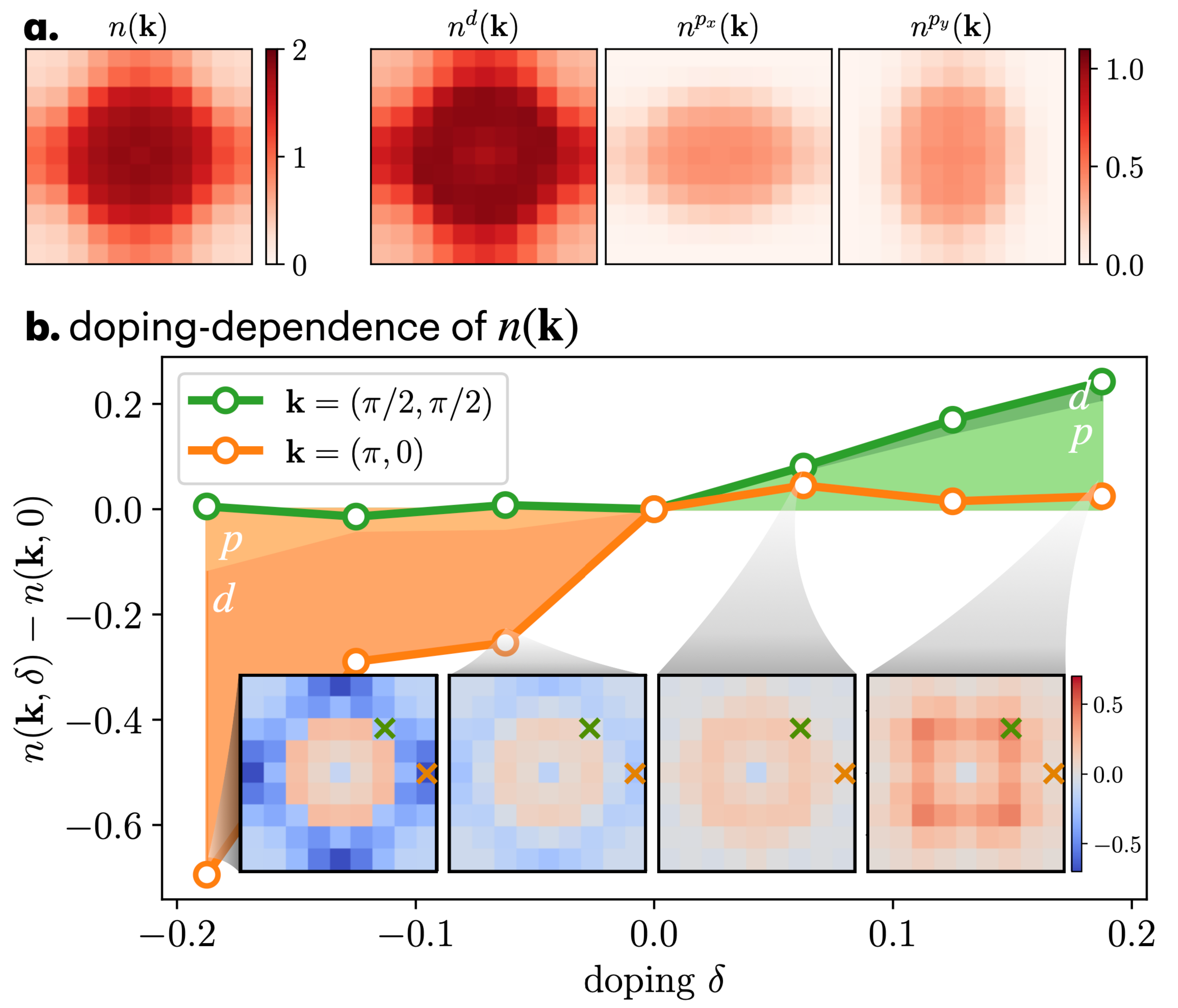}
    \caption{The density in reciprocal space, $n(\mathbf{k})$: \textbf{a.} We show the full $n(\mathbf{k})$ as well as the contributions $n^\nu(\mathbf{k})$ from $\nu=d, p_x,p_y$ (left to right) for a $10\times 10$ system with  $\delta=1/8$ and $\Delta_{pd}/t_{pd}=3.5$. \textbf{b.} The difference $n(\mathbf{k},\delta)-n(\mathbf{k},0)$ as a function of doping $\delta$ for $8\times 8$ systems. Green (orange) show the result for the nodal $\mathbf{k}=(\pi/2,\pi/2)$ (antinodal $\mathbf{k}=(\pi,0)$) and equivalent points. Insets show the full difference map for selected doping values. 
    }
    \label{fig:nk}
\end{figure}

We first consider the distribution of the doped charge in momentum space obtained from the contributions from the $d$ and $p_x,p_y$ orbitals, $n(\mathbf{k})=\sum_{\nu}n^\nu(\mathbf{k})$, with $\nu=d,\,p_x,\,p_y$ and
\begin{align}
    n^\nu(\mathbf{k})=\frac{1}{N_\mathrm{cell}}\sum_{\mathbf{i}, \mathbf{j}}e^{i\mathbf{k}\cdot (\mathbf{i}-\mathbf{j})}M(\nu,\mathbf{i}; \nu,\mathbf{j})\, ,
\end{align}
as well as the coherence matrix, 
\begin{align}
    M(\nu,\mathbf{i}; \nu^\prime,\mathbf{j})=\sum_\sigma\langle \hat{C}_{\nu\mathbf{i}\sigma}^\dagger\hat{C}_{\nu^\prime\mathbf{j}\sigma} \rangle\,.
    \label{eq:M}
\end{align}
As shown in Fig.~\ref{fig:nk}a for a $10\times 10$ system at $\delta=1/8$ hole doping, the different orbital contributions exhibit distinct momentum-space structures, with the $x$ ($y$) alignment of the $p_{x(y)}$ orbitals clearly visible also in momentum space~\footnote{As discussed earlier, due to the gauge of the Hamiltonian~\eqref{eq:3bandEmery}, the quasi-momentum $\mathbf{k}$ defined here is shifted by $\mathbf{Q}=(\pi,\pi)$ w.r.t. the Emery model with the signs from the different $p$- and $d$-orbital overlaps encoded explicitly  in the hopping signs.}.

To isolate the changes in $n(\mathbf{k})$ induced by doping, we additionally consider the difference $n(\mathbf{k},\delta)-n(\mathbf{k},0)$, shown in Fig.~\ref{fig:nk}b for $8\times8$ systems across several electron ($\delta<0$) and hole ($\delta>0$) doping values. We focus in particular on the nodal $\mathbf{k}=(\pi/2,\pi/2)$ and antinodal $\mathbf{k}=(\pi,0)$ momenta (and their symmetry-equivalent points). The doping dependence reveals a pronounced momentum-space redistribution of the density, with the nodal and antinodal regions evolving markedly differently for electron and hole doping: While for electron doping, the main contribution w.r.t. the $\delta=0$ system comes from a depletion of the $d$-signal at the antinode, the main change for hole doping occurs for the $p$-signal at the nodes. The corresponding momentum-space maps highlight this distinct evolution of nodal and antinodal regions.\\

The resulting nodal-antinodal dichotomy is qualitatively consistent with angle-resolved photoemission spectroscopy (ARPES) experiments. Although $n(\mathbf{k})$ is an equal-time quantity and therefore should not be identified directly with the ARPES spectral function, a qualitatively similar electron-hole asymmetry is observed: In hole-doped cuprates, ARPES reveals low-energy quasiparticle weight emerging first around the nodal region and forming a Fermi arc~\cite{Yoshida2003,Zhou2004}, while in electron-doped cuprates low-energy spectral weight forms an electron pocket around $(\pi,0)$~\cite{Armitage2002}. 

Furthermore, a similar phenomenology is found in effective single-band models including next-nearest neighbor hopping $t^\prime$. For $t^\prime <0 (>0)$ regimes corresponding to hole (particle) doping, exact diagonalization studies find a minimum of the single-particle dispersion at the (anti-)node~\cite{Gooding1994}. Furthermore, a gap opening near $(\pi,0)$ while the nodal region remains gapless is observed upon hole doping, leading to a Fermi-arc-like structure. For electron doping, the situation is reversed: a gap develops along the nodal direction, while low-energy weight remains concentrated near $(\pi,0)$~\cite{Tohyama2004}.

\subsubsection{Downfolding to single-band models}
Next, we systematically analyze the effect of different Emery parameter choices on effective single-band models. Denoting the single-band operators by $\hat{c}^{(\dagger)}_{\mathbf{i}\sigma}$, the most commonly studied version of the single-band Fermi-Hubbard model is 
\begin{align}
\hat{\mathcal{H}}_\mathrm{FH} = -\sum_{\langle \mathbf{ij}\rangle_r,\sigma}t(r)\left( \hat{c}_{\mathbf{i}\sigma}^\dagger \hat{c}_{\mathbf{j}\sigma}+\mathrm{h.c.}\right)+U\sum_\mathbf{i}\hat{n}_{\mathbf{i},\uparrow}\hat{n}_{\mathbf{i}\downarrow}\,,
\end{align}
with $\langle\mathbf{ij}\rangle_r$ denoting pairs at distance $r=\vert \mathbf{i}-\mathbf{j}\vert$.
In the remainder of this paper, we will often denote $t:=t(1)$, $t^\prime:=t(\sqrt{2}), t^{\prime\prime}:=t(2)$ etc. (with the same convention applying to the other $r$-dependent parameters introduced below).

Here, we also consider the presence of additional terms: $(i)$~a density-assisted hopping contribution,
\begin{align}
    \hat{\mathcal{H}}_{t_n} = -\sum_{\langle \mathbf{ij}\rangle_r}\sum_{\sigma}t_n(r)\left( \hat{c}_{\mathbf{i}\sigma}^\dagger \hat{c}_{\mathbf{j}\sigma}+\mathrm{h.c.}\right)\left(\hat{n}_{\mathbf{i}\bar{\sigma}}+\hat{n}_{\mathbf{j}\bar{\sigma}}\right),
    \label{eq:tn}
\end{align}
$(ii)$~a density interaction for $r>0$
\begin{align}
    \hat{\mathcal{H}}_V=\sum_{\langle \mathbf{ij}\rangle_r}\sum_\sigma
     V_{\sigma\bar{\sigma}}(r) \, \hat n_{i\sigma}\hat n_{j\bar\sigma} 
    \label{eq:HVupdn}
\end{align}
and $(iii)$~correlated and pair hopping terms
\begin{align}
    \hat{\mathcal{H}}_{t_c}=\sum_{\langle \mathbf{ij}\rangle_r} t_c(r)\left(\hat{c}_{\mathbf{i\uparrow}}^\dagger\hat{c}_{\mathbf{j\uparrow}} \hat{c}_{\mathbf{j}\downarrow}^\dagger\hat{c}_{\mathbf{i\downarrow}}+\hat{c}_{\mathbf{i\uparrow}}^\dagger\hat{c}_{\mathbf{j\uparrow}} \hat{c}_{\mathbf{i\downarrow}}^\dagger\hat{c}_{\mathbf{j}\downarrow}+\mathrm{h.c.}\right)\,.
    \label{eq:Htc}
\end{align}
We note that due to the nature of the downfolding-process, $t_c(r) = V_{\uparrow\downarrow}(r)$ (see Appendix~\ref{app:Wannier}). In general, also three- and four-site contributions can arise from the downfolding procedure but are neglected here due to their small magnitude. \\ 

\paragraph{Downfolding scheme.} The extended effective single-band model $\hat{\mathcal{H}}_\mathrm{1b}=\hat{\mathcal{H}}_\mathrm{FH}+\hat{\mathcal{H}}_{t_n}+\hat{\mathcal{H}}_{V}+\hat{\mathcal{H}}_{t_c}$ is derived by constructing a unitary transformation $\mathcal{W}$ such that
\begin{equation}
    \hat{c}_{\mathbf{j}\sigma}^\dagger
    =
    \sum_{\alpha}\mathcal{W}_{\mathbf{j}\alpha} \hat{C}_{\alpha\sigma}^\dagger \,,
    \label{eq:trafo}
\end{equation}
with $\alpha=(\mathbf{i},\nu)$ in the three-band model~\cite{Jiang2023}. In practice, $\mathcal{W}$ is constructed from the coherence matrix $M(\alpha;\beta)$ (see Eq.~\eqref{eq:M}) by diagonalization, which allows us to obtain the natural orbitals (the eigenvectors of $M$, $v_n(\alpha)$) and their occupation (the eigenvalues, $\lambda_n$). 

\begin{figure}[t]
    \centering
    \includegraphics[width=0.5\textwidth]{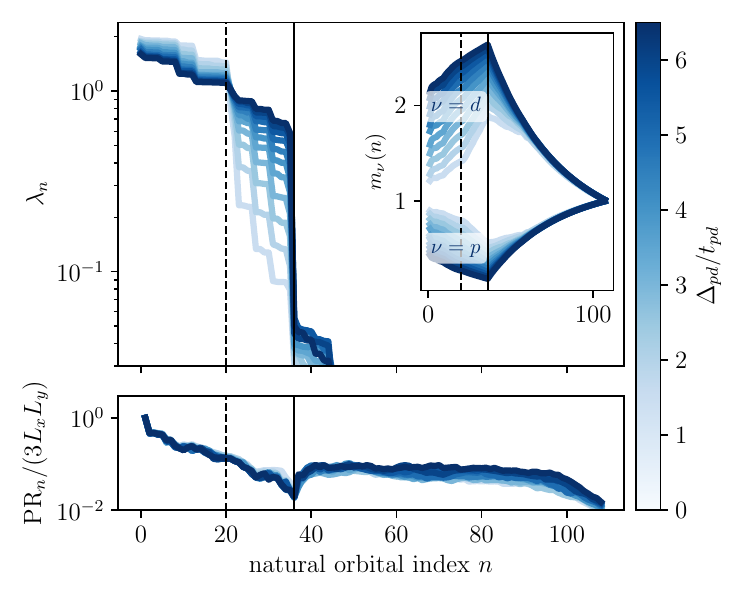}
    \caption{Natural orbitals and Wannier functions for a $6\times 6$ system at $\delta=1/8$ and with $U_d/t_{pd}=6.0$. We show the natural orbital occupation $\lambda_n$ (top panel). Solid (dashed) vertical lines correspond to $n=L_xL_y$ ($n=N_h/2$). The inset displays the relative orbital mass $m_\nu(n)$ when occupying the corresponding natural orbitals $v_n$ up to the index $n$, see Eq.~\eqref{eq:mass}. Wannier functions $w_n$ are constructed as superpositions of $v_n$ projected onto the subspace $\{v_l\}_{l<n}$. Small values of the participation ratio $\mathrm{PR}_n$, Eq.~\eqref{eq:PR}, indicate a high degree of localization.
    }
    \label{fig:EVs}
\end{figure}

\begin{figure*}[t]
    \centering
    \includegraphics[width=0.98\textwidth]{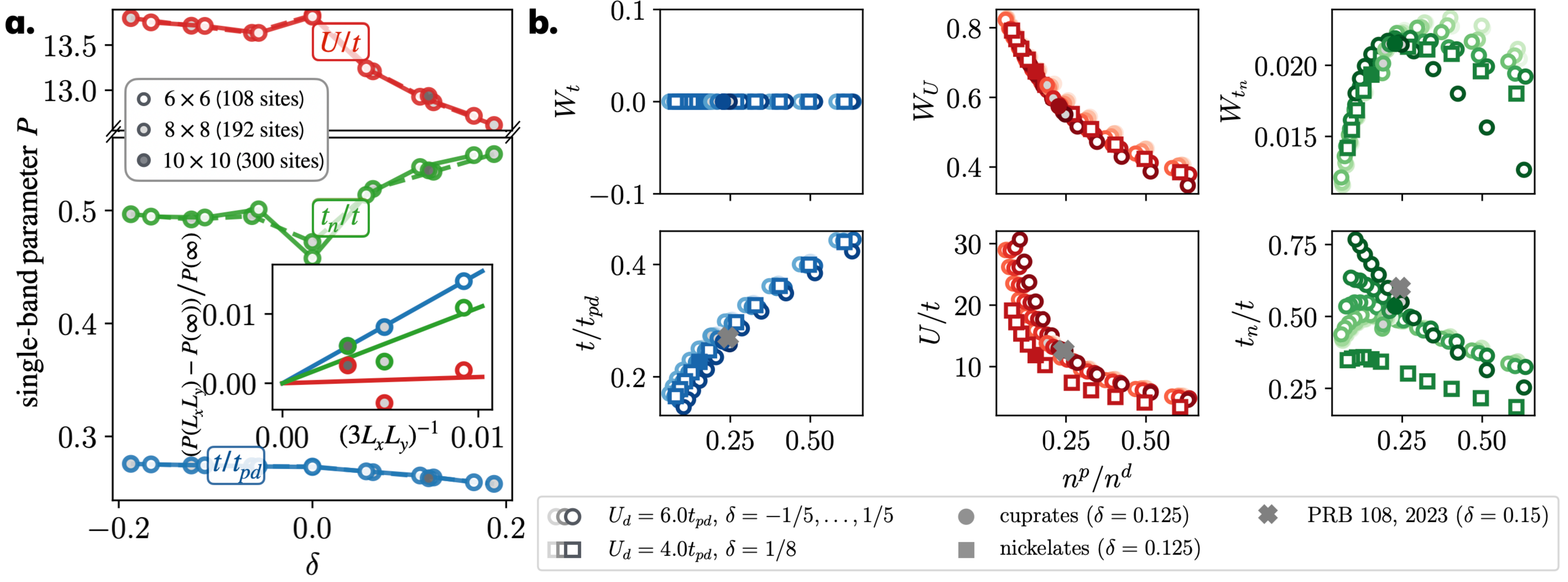}
    \caption{\textbf{a.} The largest contributions to the downfolded single-band model as a function of electron ($\delta<0$ and hole ($\delta>0$) doping for $U_d/t_{pd}=6.0$ and $\Delta_{pd}/t_{pd}=3.5$: nearest-neighbor tunneling $t$ (blue), on-site repulsion $U$ (red) and density-assisted hopping $t_n$ (green), see Eq.~\eqref{eq:tn}. 
    Markers with white, light gray and dark gray filling indicate the result for $6\times 6$, $8\times 8$ and $10\times 10$ unit cell systems, respectively. The inset compares the effect of different system sizes explicitly by showing the difference for each parameter and each system size, $P(L_xL_y)$, w.r.t. the extrapolated value, $P(\infty)$. \textbf{b.} The largest contributions to the downfolded single-band model as a function of the relative $p$- vs. $d$-site densities, $n^p/n^d$. We plot different data groups: the cuprate- (nickelate-) inspired regime with $U_d/t_{pd}=6.0\, (4.0)$, varying the charge-transfer energy $\Delta_{pd}/t_{pd}$ and doping $\delta$, with the latter denoted by light to dark colors. We mark the reconstructions for the simulation most closely to the usually considered parameters and $\delta=1/8$ with filled colored markers. The reconstruction obtained in Ref.~\cite{Jiang2020criticalNatureoftheNiSpinState} for a cylinder and $\delta=0.15$ is indicated by the gray cross.}
    \label{fig:params}
\end{figure*}

The natural orbital occupancy $\lambda_n$ for a $6\times 6$ unit cell system at fixed $\delta=1/8$ and different values of $\Delta_{pd}/t_{pd}$ are shown in Fig.~\ref{fig:EVs}a. There is a sharp drop in $\lambda_n$ at $n=L_xL_y$ (solid black line) for intermediate to large $\Delta_{pd}/t_{pd}\geq 3$, while for smaller $\Delta_{pd}/t_{pd}$ the drop occurs at $n=N_h/2$ with $N_h=40$ for the $\delta=1/8$ considered here (dashed black line). To gain insight into the origin of these different behaviors, we calculate the relative weight of orbital $\nu=d,p_x,p_y$ when occupying the natural orbitals $v_n(\alpha=\mathbf{i},\nu)$ up to the index $n$,
\begin{align}
    m_\nu(n) = \sum_{l=0}^{n-1}V_{l,\nu}/\sum_{ \nu^\prime}\sum_{l^\prime=0}^{n-1} V_{l^\prime,\nu^\prime}\,,
    \label{eq:mass}
\end{align}
with $V_{n,\nu}=\sum_{\mathbf{i}} \vert v_{n}(\mathbf{i},\nu)\vert^2$.
This is shown in the inset of Fig.~\ref{fig:EVs}a. It can be seen that while for intermediate-to large $\Delta_{pd}/t_{pd}\geq 3$ natural orbitals with a large $d$-mass are occupied, for small $\Delta_{pd}/t_{pd}$ comparably more $p$-mass is present until $n=N_h/2$. For these orbitals, the effective cost of placing two holes in one orbital is lower since $U_p<U_d$. Correspondingly, these occupations are close to $\lambda \approx 2$ and can be filled up to $n=N_h/2$, where the drop is observed.\\

To obtain  the transformation matrix $\mathcal{W}$, the natural orbitals are localized and orthonormalized in order to construct Wannier functions  $w_n(\alpha)$, see Fig.~\ref{fig:1}c. Due to the large $d$-weight at intermediate-to large $\Delta_{pd}/t_{pd}\geq 3$, we expect Wannier functions centered at the $d$-sites. Therefore, we can construct them as in Ref.~\cite{Jiang2023} by centering the active $L_xL_y$ natural orbitals explicitly on these sites. In contrast, for smaller $\Delta_{pd}/t_{pd}$ this is not as clear due to the larger $p$-weight. Hence, we use a more general scheme that constructs the Wannier functions as the eigenstates of the position operator projected to the subspace of $n$ natural orbitals, see also Appendix~\ref{app:Wannier}. To characterize the resulting Wannier orbitals, we compute the participation ratio,
\begin{equation}
\mathrm{PR}_n
=\frac{1}{n}\sum_{l=0}^{n-1}
\frac{1}
{\sum_{\alpha} \vert w_l(\alpha)\vert ^4},
\label{eq:PR}
\end{equation}
which estimates the localization of the Wannier functions. We find that $\mathrm{PR}_n$ is minimized for $n=L_xL_y$ even for small $\Delta_{pd}/t_{pd}$. Hence, it is best to localize on the $d$-sites using $N_\mathrm{act}=L_xL_y$ in all cases. We find it convenient to replace the index $n$ by the position of the Wannier function centers, $w_{\mathbf{i}_d}(\alpha)$ with $\mathbf{i}_d$ referring to $d$-sites from now on.
In this notation, the effective parameters are arise from Wannier function overlaps of the form \begin{equation}
    \begin{aligned}
    t(r)&=-\frac{1}{N_{r}}\sum_{\langle \mathbf{i}_d,\mathbf{j}_d\rangle_r} \sum_{\alpha,\beta}P_{\alpha\beta}\,w_{\mathbf{i}_d}(\alpha)w_{\mathbf{j}_d}(\beta), \\    U&=\frac{1}{N_\mathrm{cell}}\sum_{\mathbf{i}_d}\sum_\alpha U_\alpha w_{\mathbf{i}_d}(\alpha)^4,\\ t_n(r)&=-\frac{1}{{N_\mathrm{r}}}\sum_{\langle \mathbf{i}_d,\mathbf{j}_d\rangle_r} \sum_\alpha U_\alpha w_{\mathbf{i}_d}(\alpha)^3w_{\mathbf{j}_d}(\alpha), 
    \label{eq:Paramformulas}
\end{aligned}
\end{equation}
for $N_r$ terms in the sum over distance-$r$ $d$-sites, $\langle \mathbf{i}_d,\mathbf{j}_d\rangle_r$, and $\alpha,\beta$ denoting all sites that are connected by the Emery parameters $P_{\alpha\beta}=t_{pd},t_{pp},\Delta_{pd}$. 
\\

\paragraph{Downfolding results.}
The results of the downfolding procedure are summarized in Fig.~\ref{fig:params}. Fig.~\ref{fig:params}a shows the doping dependence of the dominant single-band parameters for both electron ($\delta<0$) and hole ($\delta>0$) doping: the nearest-neighbor hopping $t$ (blue), the on-site interaction $U$ (red), and the density-assisted hopping $t_n$ (green). Markers with white, light-gray, and dark-gray filling denote results obtained on $6\times6$, $8\times8$, and $10\times10$ unit-cell systems, respectively, with the largest system size included only at $\delta=1/8$. The data for the different system sizes are nearly indistinguishable, indicating that finite-size effects are small. This is illustrated more explicitly in the inset, which shows the deviation of each parameter $P=t,U,t_n$ from its extrapolated thermodynamic-limit value, $P(L_xL_y)-P(\infty)$. The weak system-size dependence motivates us to restrict the analysis to $6\times6$ systems in some of the subsequent discussion.\\

The doping dependence of the effective parameters exhibits a pronounced asymmetry between electron and hole doping. While the effective parameters remain comparatively weakly varying over the investigated electron-doped regime, the ratio $U/t$ is largest at half filling and decreases monotonically upon hole doping. At the same time, $t_n/t$ reaches a minimum at $\delta=0$ and increases with hole doping. 

In order to understand the evolution of the effective parameters we take a look at Fig.~\ref{fig:params}b, which collects several datasets simulated using HFDS. We distinguish between a cuprate-inspired regime, characterized by $U_d/t_{pd}=6.0$ and simulated for several $\delta=-1/5,\dots,1/5$, and a nickelate-inspired regime with $U_d/t_{pd}=4.0$ and simulated for $\delta=1/8$. We put particular emphasis on varying the charge-transfer energy $\Delta_{pd}/t_{pd}$, representing a tuning knob of the charge distribution between different orbitals, $n^p/n^d$, which has been shown to affect the critical temperatures in several works~\cite{ocallaghan2026chargetransfergapsizeoxygen,Weber_2012,Ruan2016Relationshipbetweentheparentchargetransfergap,Wang2023Correlatingthecharge-transfergaptothemaximumtransitiontemperature,Yee2014charge-transferenergyinhole-dopedcuprates}. The filled square and round markers indicate the reconstructions corresponding to the parameter sets used in our simulations and $\delta=1/8$, for the cuprate- and nickelate-inspired regimes, respectively. For comparison, the gray cross shows the reconstruction reported in Ref.~\cite{Jiang2023} for a cylindrical geometry at $\delta=0.15$~\footnote{Since the work~\cite{Jiang2023} only provides the doped hole densities and not the full densities, we estimate $n^p/n^d$ from our values obtained at half-filling and add the doped hole densities provided in Ref.~\cite{Jiang2023}}. 

Fig.~\ref{fig:params}b displays the reconstructed single-band parameters $t,U,t_n$ (lower panel) as a function of the relative $p$- vs. $d$-site occupancy, $n^p/n^d$, for all parameter sets. Furthermore, we show their respective Wannier overlaps (upper panel) given by 
\begin{equation}
    \begin{aligned}
        W_t&=\frac{1}{N_r}\sum_{\langle\mathbf{i}_d,\mathbf{j}_d\rangle_r}\sum_{\alpha} w_{\mathbf{i}_d}(\alpha)w_{\mathbf{j}_d}(\alpha)\,,\\
        W_U&=\frac{1}{N_\mathrm{cell}}\sum_{\mathbf{i}_d}\sum_\alpha  w_{\mathbf{i}_d}(\alpha)^4,\\ 
        W_{t_n}&=\frac{1}{N_r}\sum_{\langle\mathbf{i}_d,\mathbf{j}_d\rangle_r}\sum_\alpha w_{\mathbf{i}_d}(\alpha)^3w_{\mathbf{j}_d}(\alpha)\,.
        \label{eq:Ws}
    \end{aligned}
\end{equation} 
Since the Wannier functions are orthogonal, $W_t=0$ up to numerical errors, see Fig.~\ref{fig:params}b (upper left panel). Increasing $n^p/n^d$ corresponds to more delocalized Wannier orbitals $w_{\mathbf{i}_d}(\alpha)$. Hence, we observe increasing effective hopping $t/t_{pd}$ and decreasing effective interaction $U/t_{pd}$ w.r.t. $n^p/n^d$ (see left and middle panels). For $t_n/t$ (right panel) the behavior is non-monotonic and depends heavily on the doping.  

Fig.~\ref{fig:params}b reveals three tuning knobs of the effective parameters: The first is the fraction $n^p/n^d$, which is reflected in drastic changes of the Wannier orbital contributions, $W_{t,U,t_n}$, and can be tuned e.g. by the charge-transfer energy $\Delta_{pd}/t_{pd}$. This can be seen in the upper panel of Fig.~\ref{fig:params}b: The Wannier contributions $W_{t,U,t_n}$ from different data sets collapse to the same behavior w.r.t. $n^p/n^d$ (except for $t_n$ at large $n^p/n^d$, where the evolution of $W_{t_n}$ is more complicated). We have confirmed that the same trends are also reflected in the absolute $U/t_{pd}$ and $t_n/t_{pd}$ (see Appendix~\ref{appendix:adddownfolding}). The second tuning knob is doping $\delta$ (indicated by the light to dark colors). As can be seen from Fig.~\ref{fig:params}b, doping leaves the Wannier contributions relatively unchanged (top panel), but leads to variations in $t/t_{pd}$ for small $n^p/n^d$ (bottom panel), which in turn affects the relative quantities $U/t$ and $t_n/t$. The third tuning knob are other parameters like $U_d/U_p$ that tune the ratio of the Wannier contributions to $U$ and $t_n$ from $d$ and $p$ sites, leading to distinctly different results for the nickelate parameter set (squares in the lower panels). To conclude, our analysis demonstrates that the effective parameters have very distinct dependence on the different tuning knobs given by the charge transfer energy, the doping and other interaction parameters.\\

\begin{table*}[]
\begin{tabular}{|c||c||c||c|c|c|c|c|c|c|c|c|}
\hline
                            & Emery parameters $[t_{pd}]$       & $n^p$ &$t/t_{pd}$ & $U/t$ & $t_n/t$ & $t^\prime/t$ &$t_n^\prime/t$  &  $t^{\prime\prime}/t$&  $t_n^{\prime\prime}/t$ & $V_{\uparrow\downarrow}/t$&$V_{\sigma\sigma}/t$\\\hline\hline
\multirow{2}{*}{cuprates}   & $U_d=6.0$, $U_p=3.0$, $\Delta_{pd}=3.5$, $t^\prime_{pp}=0.0$, $V_{pd}=0.0$& 0.17 & 0.27 & 12.93 & 0.54 & 0.07 & 0.05 & -0.05 & -0.07 & 0.09 &-                            \\
& $U_d=8.0$, $U_p=3.5$, $\Delta_{pd}=3.0$, $t^\prime_{pp}=0.0$, $V_{pd}=0.5$ &0.19 & 0.29 & 15.24 & 0.59 & 0.06 & 0.05 & -0.06 & -0.1 & 0.29$^{*}$ &0.18\\
& $U_d=6.0$, $U_p=3.0$, $\Delta_{pd}=3.5$, $t^\prime_{pp}=0.15$, $V_{pd}=0.0$ &                        
                            0.19 & 0.3 & 11.3 & 0.54 & 0.11 & 0.07 & -0.02 & -0.05 & 0.1&-
                             \\
& $U_d=6.0$, $U_p=3.0$, $\Delta_{pd}=1.5$, $t^\prime_{pp}=0.15$, $V_{pd}=0.0$ &0.29 & 0.46 & 5.18 & 0.38 & 0.07 & -0.01 & -0.04 & -0.06 & 0.12&- \\\hline\hline
\multirow{2}{*}{nickelates} & $U_d=4.0$, $U_p=3.0$, $\Delta_{pd}=4.5$, $t^\prime_{pp}=0.0$, $V_{pd}=0.0$ &0.13 & 0.23 & 11.87 & 0.35 & 0.08 & 0.04 & -0.03 & -0.04 & 0.05&- \\
                            & $U_d=6.0$, $U_p=3.0$, $\Delta_{pd}=6.5$, $t^\prime_{pp}=0.0$, $V_{pd}=0.0$& 0.08 & 0.16 & 29.36 & 0.63 & 0.05 & 0.13 & -0.03 & -0.02 & 0.05&-\\\hline
\end{tabular}
\caption{Downfolded single-band parameters and $p$-site density, $n^p$, for $\delta=1/8$ and some selected Emery parameter choices typically considered for cuprates and nickelates. $^{*}$For $V_{pd}\neq 0$, $t_c\neq V_{\uparrow\downarrow}$ (see Appendix~\ref{app:Wannier}). In row 2, we find $t_c/t=0.11$. }
\label{tab:params}
\end{table*}

Lastly, we present downfolding results for selected sets of Emery parameters commonly considered for cuprates and nickelates in Tab.~\ref{tab:params}. In most cases, we find $U$ significantly larger than the typical $U/t=8$. However, taking into account the effective hopping generated by $t_n$ with the mean density $\langle n\rangle$ in the single band, $U/(t+t_n\langle n\rangle )\approx 8$~\footnote{In contrast, the effective spin exchange obtained via Schrieffer-Wolff transformation is directly enhanced $J\propto (t+t_n)^2/U$.}. For a more detailed discussion, we begin with a canonical parameter set~\cite{Jiang2023} that is used as a starting point throughout this work (line 1). In line 2, we consider a set including a repulsion $V_{pd}\,\hat{n}^p_{\mathbf{i}\uparrow} \hat{n}^d_{\mathbf{j}\downarrow}$ between neighboring $p$ and $d$ sites as employed in several works~\cite{Hanke2010,Cui2020,White2015}. While $n^p$ and the largest effective single-band parameters remain very similar between these cases, including $V_{pd}$ in the downfolding significantly enhances $V_{\uparrow\downarrow}$ and moreover generates additional equal-spin density terms in analogy to Eqs.~\eqref{eq:HVupdn} with amplitude $V_{\sigma{\sigma}}(r)$ that are not present for $V_{pd}=0$. Furthermore, Ref.~\cite{jacob2026conventionalemerymodelcrucial} discusses the importance of long-range hoppings; accordingly, we also include $t_{pp}^\prime$ in lines 3 and 4.
For nickelates, $\Delta_{pd}>U_d$ and we consider one value of $U_d$ at the lower edge of the range of realistic parameters discussed in the literature, as well as a larger value. We would like to point out that in all settings considered in this work, we find a significant $t_n$, supporting the observation that was made in Ref.~\cite{Jiang2023} for selected parameters. In addition, the effective ratio $t^\prime/t$ is smaller than the values typically considered~\cite{Xu2024,rende2026superconductivitytthubbardmodel}. The electron-hole asymmetry is therefore captured by $t_n$, rather than by assigning different signs to $t^\prime$.

\section{Discussion}
Our results demonstrate that NQS, in particular HFDS, provide an efficient variational description of the copper- and nickel-oxide layers of cuprate and nickelate superconductors that goes beyond the usually considered single-band Fermi-Hubbard model: the three-band Emery model. We demonstrate that the favorable scaling of HFDS enables accurate calculations on square lattices of up to $10\times10$ unit cells, establishing NQS as a practical route to studying the ground state of the Emery model on large two-dimensional systems. As an example, our simulations reveal a clear asymmetry between electron and hole doping in momentum space, which is qualitatively consistent with ARPES experiments in cuprates. 

The central result of this work is the systematic downfolding of the correlated three-band ground state to an effective single-band description. We find sizable contributions beyond the bare Fermi-Hubbard model as in Ref.~\cite{Jiang2023}. Using HFDS allows us to extend the analysis to a large variety of Emery model parameters, including different system sizes, hole and particle dopings, varying charge transfer energies and other model parameters relevant to cuprates and nickelates. The results of this analysis can be summarized as follows: $(i)$ the effective parameters show a doping dependence; $(ii)$ the charge-transfer energy, previously shown to have a strong impact on the critical temperatures~\cite{ocallaghan2026chargetransfergapsizeoxygen,Weber_2012,Ruan2016Relationshipbetweentheparentchargetransfergap,Wang2023Correlatingthecharge-transfergaptothemaximumtransitiontemperature,Yee2014charge-transferenergyinhole-dopedcuprates}, has a particularly strong impact on the spatial extent of the interacting Wannier orbitals, Eq.~\eqref{eq:Ws}, and, consequently, of the effective hopping and interaction strengths given by Eq.~\eqref{eq:Paramformulas}; and $(iii)$ interaction-generated terms beyond the conventional Hubbard model, most notably density-assisted hopping, remain sizable throughout all parameter regimes considered here, with a stronger contribution to the particle-hole asymmetry than next-nearest neighbor hoppings.

Our results highlight the potential of NQS as a tool for studying multi-band strongly correlated systems. Looking ahead, the favorable scaling of these methods will allow us not only to derive the more accurate effective single-band models that were obtained in this work, but also to investigate the properties of the Emery model itself in greater detail, including its superconducting properties. This provides a promising route toward establishing a direct connection between microscopic multi-band models and the low-energy physics of cuprate and nickelate superconductors.

\section{Methods}
\label{sec:methods}

Throughout this paper, we employ NQS~\cite{Carleo2017,lange2024reviewnqs,Medvidovic2024}, in particular HFDS~\cite{Moreno2022}. For width-two systems, where tensor-network calculations are feasible, we additionally benchmark our HFDS results against MPS calculations. We introduce the two methods below.\\

\textit{Hidden Fermion Determinant States (HFDS).---} HFDS are motivated by the determinant representation of free-fermion wave functions in second quantization. 
In a Fock basis configuration
$\ket{\mathbf{\sigma}}
=
\hat{c}_{\mathbf{i}_0}^\dagger\cdots\hat{c}_{\mathbf{i}_{3N_\mathrm{cell}}}^\dagger\ket{0}$,
the corresponding wave-function amplitude is given by a Slater determinant,
\begin{align}
\langle\mathbf{\sigma}|\psi_0\rangle
=
\det(\mathbf{\sigma}\star\mathbf{U}),
\end{align}
where $\mathbf{\sigma}\star\mathbf{U}$ selects the rows of $\mathbf{U}$ corresponding to the occupied single-particle states~\cite{Becca_Sorella_2017}.

HFDS extend this construction by augmenting the Hilbert space with auxiliary, or hidden, fermionic degrees of freedom and subsequently projecting onto the physical subspace~\cite{Moreno2022}. For $N_v$ visible and $N_h$ hidden fermions, the resulting amplitudes are given by Eq.~\eqref{eq:HFDS},
where the hidden-fermion configuration depends on the visible configuration and the configuration-dependent matrices $\mathbf{U}_{vh,2}(\mathbf{\sigma})$ and $\mathbf{U}_h(\mathbf{\sigma})$ are parametrized by a neural network~\cite{Chen2025Thesis,chen2025neuralnetworkaugmentedpfaffianwavefunctions}. HFDS can equivalently be expressed as a neural backflow transformation with a specific form for the Jastrow factor~\cite{Liu2024unifying}, consistent with numerical observations that the two ansätze exhibit comparable performance~\cite{Liu2024unifying,sharma2025comparingsymmetrizeddeterminantneural}.\\

In our calculations, we use $N_h=10$ hidden fermions and parameterize the configuration dependence of the lower, hidden block in Eq.~\eqref{eq:HFDS} using convolutional neural networks (CNNs) with $n_l=3$ layers, $f=60$ features per layer, and kernel size $k=3$. This results in approximately $2\times10^5$ variational parameters. We enforce total-spin conservation and, for systems larger than $6\times6$ unit cells, symmetry under translations by half the system size through symmetry-averaging~\cite{chen2025NeuralnetworkaugmentedPfaffian}. More extensive benchmarks and hyperparameter scans of this ansatz can be found in Appendix~\ref{appendix:benchmarks}.\\

\textit{Matrix Product States (MPS).---} For width-two systems, where MPS calculations are computationally feasible, we use the MPS implementation in SyTen~\cite{syten1,syten2} with a maximum bond dimension of $\chi_{\mathrm{max}}=4096$. We perform both single-site and two-site density-matrix renormalization group (DMRG) calculations and enforce particle-number and spin conservation throughout.

\section*{Acknowledgments} 
We wish to thank Antoine Georges, Chris Roth and Shengtao Jiang for valuable discussions. This research was supported by the Deutsche
Forschungsgemeinschaft (DFG, German Research Foundation)
under Germany’s Excellence Strategy EXC-2111 Grant
No. 390814868 and the European Research Council
(ERC) under the European Union’s Horizon 2020 research
and innovation program (Grant Agreement No. 948141),
ERC Starting Grant SimUcQuam and ERC Starting Grant QuaQuaMA (Grant No. 101217531). Numerical simulations were performed on the Paderborn and Arnold Sommerfeld Cluster. The HFDS simulations are performed using Quantax \cite{quantax} and lrux \cite{chen2026lruxfastlowrankupdates}. DMRG calculations have been done using the \textsc{SyTen} toolkit, developed and maintained by C. Hubig, F. Lachenmaier, N.-O. Linden, T. Reinhard, L. Stenzel, A. Swoboda, M. Grunder, S. Mardazad, F. Pauw and S. Paeckel. Information is available at \href{https://syten.eu/}{www.syten.eu}.





\bibliography{references.bib}
\widetext


\newpage
\pagebreak
\appendix

\newpage

\newpage
\pagebreak
\appendix
\widetext

\newpage

\begin{center}
\textbf{\large Appendix}

\end{center}
\setcounter{equation}{0}
\setcounter{figure}{0}
\setcounter{table}{0}
\setcounter{page}{1}
\makeatletter
\renewcommand{\theequation}{S\arabic{equation}}
\renewcommand{\thefigure}{S\arabic{figure}}

\section{Numerical Downfolding Protocol}
\label{app:Wannier}

Starting from the Emery model with one copper $d$ orbital and two oxygen
$p_x$ and $p_y$ orbitals, we construct Wannier functions localized on the
copper sites. As a first step, we determine the coherence matrix 
\begin{align}
M_{\mathbf{i},\mathbf{j}} =
\begin{pmatrix}
M(d,\mathbf{i};d,\mathbf{j}) & M(d,\mathbf{i};p_x,\mathbf{j}) & M(d,\mathbf{i};p_y,\mathbf{j}) \\
M(p_x,\mathbf{i};d,\mathbf{j}) & M(p_x,\mathbf{i};p_x,\mathbf{j}) & M(p_x,\mathbf{i};p_y,\mathbf{j}) \\
M(p_y,\mathbf{i};d,\mathbf{j}) & M(p_y,\mathbf{i};p_x,\mathbf{j}) & M(p_y,\mathbf{i};p_y,\mathbf{j})
\end{pmatrix},
\end{align}
with $M(\nu,\mathbf{i};\nu^\prime,\mathbf{j}) $ given by Eq.~\eqref{eq:M}. We furthermore restore translational invariance by averaging this matrix over all lattice translations,
\begin{equation}
\bar M_{\mathbf{ij}}
=
\frac{1}{N_T}
\sum_{\mathbf R}
M_{\mathbf{i}+\mathbf R,\mathbf{j}+\mathbf R},
\end{equation}
where the sum runs over the set of lattice translation vectors \(\mathbf R\), and \(N_T\) is the number of translations considered.

The translation-averaged coherence matrix is diagonalized,
\begin{equation}
\bar M
=
V\Lambda V^\dagger,
\end{equation}

with eigenvalues ordered as $\lambda_0\ge\lambda_1\ge\cdots$. The first \(N_{\rm act}\) natural orbitals define the active subspace,
\begin{equation}
U=\left(
v_0(\alpha),v_1(\alpha),\ldots,v_{N_{\rm act}-1}(\alpha)
\right),
\end{equation}
where $\alpha=(\nu,\mathbf{i})$.
In order to construct the transformation to the effective single-band basis, the natural orbitals are localized and orthonormalized. This results in localized Wannier functions $w_n(\alpha)$ (also labeled according to the respective Wannier center $\mathbf{i}$ in the main text, $w_{\mathbf{i}}(\alpha)$), according to which a matrix $\mathcal{W}=(w_{\mathbf{i}_0}(\alpha), \dots, w_{\mathbf{i}_{N_\mathrm{act}}}(\alpha))^T$ and the transformation
\begin{equation}
    \hat{c}_\mathbf{j}^\dagger
    =
    \sum_{\alpha}w_\mathbf{j}(\alpha) \hat{C}_\alpha^\dagger ,
    \label{eq:trafo_SM}
\end{equation}
is constructed.

\subsection{Wannier function construction} 

Below, we explain two protocols to obtain the localized Wannier functions $w_\mathbf{i}(\alpha)$. 

\paragraph{Wannier function Centered at Copper Sites}
To explicitly center the natural orbitals orbitals on the copper sites, we follow Ref.~\cite{Jiang2023} and construct
\begin{equation}
    \psi_{(d,\mathbf{i})}(\alpha) = \sum_n v_n(d,\mathbf{i})\, v_n(\alpha),
\end{equation}
where $(d,\mathbf{i})$ denotes the position vector of a copper $d$-site at unit cell $\mathbf{i}$.

\paragraph{General Wannier functions}
Let \(\{|\alpha\rangle\}\) denote the microscopic orbital basis, where each
orbital \(\alpha\) is assigned its real-space coordinate, $\mathbf r_\alpha=(x_\alpha,y_\alpha)$, where Cu orbitals occupy integer lattice sites $(x,y)$,  while oxygen orbitals are located halfway along the Cu-O bonds, $(x+\tfrac12,y)$ and $(x,y+\tfrac12)$ respectively.
The position operators are diagonal in this basis,
\begin{equation}
X_{\alpha\beta}=\sum_\alpha x_\alpha \ket{\alpha} \bra{\alpha} ,
\qquad
Y_{\alpha\beta}=\sum_\alpha y_\alpha \ket{\alpha} \bra{\alpha}.
\end{equation}
These operators $X$ and $Y$ are projected into the active subspace,
\begin{equation}
X_P=U^\dagger XU,
\qquad
Y_P=U^\dagger YU.
\end{equation}
We obtain localized orbitals by diagonalizing the projected position operator~\cite{Kivelson1982},
\begin{equation}
R_P=X_P+iY_P.
\end{equation}
Without the projection, the corresponding eigenvalues are given by the three-band positions,
$R\ket{\alpha}
=
(x_\alpha+i y_\alpha)\ket{\alpha}$ with $X \ket{\alpha}=x_\alpha\ket{\alpha}$ and $Y\ket{\alpha}=y_\alpha\ket{\alpha}$ fulfilled individually. After projection, eigenstates of $R_P$, denoted as $\psi_z(\alpha)$, fulfill
\[
R_P|\psi_z(\alpha)\rangle=z|\psi_z(\alpha)\rangle,
\]
i.e. the combined complex \(z=x+iy\) plays the role of the effective coordinate associated with the
corresponding orbital, $x$ and $y$ not not necessarily correspond to three-band positions. Within the projected subspace, the $\psi_z(\alpha)$ are hence the eigenfunctions that are maximally distinguished by their position $z$.\\

In both cases, the resulting $\psi_\mathbf{i}(\alpha)$ are not necessarily orthogonal. We therefore perform a Löwdin symmetric orthogonalization. Defining the overlap matrix $S=\psi^\dagger\psi $, the final orthonormal Wannier basis is
\begin{equation}
\mathcal{W}
=
\psi S^{-1/2}.
\end{equation}

\subsection{The transformation}
The columns of the matrix $\mathcal{W}$ are constructed from the orthonormalized
copper-centered Wannier orbitals, $\mathcal{W}=\{w_{\mathbf{i}_d}(\alpha)\}$.
The effective Hamiltonian parameters are obtained by projecting the original
three-band Emery Hamiltonian onto this Wannier basis according to Eq.~\eqref{eq:trafo}. This gives rise to the following effective single-band contributions.
\begin{itemize}
    \item Hopping contributions are determined from
    \begin{align}
    t(d)=-\frac{1}{N_{r}}\sum_{\langle \mathbf{i}_d,\mathbf{j}_d\rangle_r} \sum_{\alpha,\beta}P_{\alpha\beta}\,w_{\mathbf{i}_d}(\alpha)w_{\mathbf{j}_d}(\beta),
\end{align}
for $N_r$ terms in the sum over distance-$r$ $d$-sites $\langle \mathbf{i}_d,\mathbf{j}_d\rangle_r$ and $\alpha,\beta$ denoting all sites that are connected by the Emery parameters $P_{\alpha\beta}=t_{pd},t_{pp},\Delta_{pd}$.
    \item Similarly, the effective on-site interaction is obtained from the copper
weight of the Wannier orbital,
\begin{equation}
U=\frac{1}{N_\mathrm{cell}}\sum_{\mathbf{i}_d}\sum_\alpha U_\alpha w_{\mathbf{i}_d}(\alpha)^4.
\label{eq:effU}
\end{equation}
This expression might imply a dependence of $U/t_{pd}$ on $U_\alpha$ rather than on $\Delta_{pd}$, as expected e.g. from Refs.~\cite{Hybertsen1992,Zheng2018FromRealMaterialstoModelHamiltonians}. However, we show in Appendix~\ref{appendix:adddownfolding} that $\frac{1}{N_\mathrm{cell}}\sum_{\mathbf{i}_d}\sum_\alpha  w_{\mathbf{i}_d}(\alpha)^4\propto \Delta_{pd}/U_\alpha$ for a wide range of parameters, resulting in an overall scaling of $U/t_{pd}\propto \Delta_{pd}$.
\item The nearest-neighbor density-assisted tunneling term is obtained from,
\begin{equation}
t_n(r)=-\frac{1}{N_\mathrm{r}}\sum_{\langle \mathbf{i}_d,\mathbf{j}_d\rangle_r} \sum_\alpha U_\alpha w_{\mathbf{i}_d}(\alpha)^3w_{\mathbf{j}_d}(\alpha).
\end{equation}

\item The nearest-neighbor density-density interaction, correlated hopping and pair-exchange acquire the same prefactor, as they all arise from the following Wannier-overlaps:
\begin{equation}
t_c(r) = V_{\uparrow\downarrow}(r)=\frac{1}{N_\mathrm{r}}\sum_{\langle \mathbf{i}_d,\mathbf{j}_d\rangle_r} \sum_\alpha U_\alpha w_{\mathbf{i}_d}(\alpha)^2 w_{\mathbf{j}_d}(\alpha)^2.
\label{eq:V_updown-downfolded}
\end{equation}
\end{itemize}

\textit{Including nearest-neighbor density interactions:} In the spirit of \cite{Hanke2010,Cui2020,White2015}, we additionally implemented a density interaction in the three-band model between nearest-neighbor $p$- and $d$-sites, as included in Table \ref{tab:params}. The term adds to Eq. (\ref{eq:3bandEmery}) 
\begin{equation}
    \hat{\mathcal{H}}_{\mathrm{3b}+V_{dp}} = \hat{\mathcal{H}}_\mathrm{3b}+\frac{1}
    {2} \sum_{\nu=p_x, p_y}\sum_{ \mathbf{i}\mathbf{j} , \sigma}V_{dp}^{\mathbf{i}\mathbf{j}}\hat n^\nu_\mathbf{i}\hat n^d_\mathbf{j}
\end{equation}
with $\hat n^\nu=\hat n^\nu_\uparrow+ \hat n^\nu_\downarrow$. Specifically, we consider a nearest-neighbor $V_{dp}$ interaction, which modifies the other single-band parameters $P$ by $\delta P$:
\begin{equation}
    \begin{aligned}
    \delta U&=\frac{2V_{pd}}{N_\mathrm{cell}}\sum_{\mathbf{i}_d}\sum_{\langle\alpha\beta\rangle_{pd}} w_{\mathbf{i}_d}(\alpha)^2\,w_{\mathbf{i}_d}(\beta)^2,\\
    \delta t_n(r)&=-\frac{V_{pd}}{N_r}\sum_{\langle \mathbf{i}_d,\mathbf{j}_d\rangle_r}\sum_{\langle\alpha\beta\rangle_{pd}}\Big[w_{\mathbf{i}_d}(\alpha)^2\,w_{\mathbf{i}_d}(\beta)w_{\mathbf{j}_d}(\beta)+w_{\mathbf{i}_d}(\beta)^2\,w_{\mathbf{i}_d}(\alpha)w_{\mathbf{j}_d}(\alpha)\Big],\\
    \delta V_{\uparrow\downarrow}(r)&=\frac{V_{pd}}{N_r}\sum_{\langle \mathbf{i}_d,\mathbf{j}_d\rangle_r}\sum_{\langle\alpha\beta\rangle_{pd}}\Big[w_{\mathbf{i}_d}(\alpha)^2\,w_{\mathbf{j}_d}(\beta)^2+w_{\mathbf{i}_d}(\beta)^2\,w_{\mathbf{j}_d}(\alpha)^2\Big],\\
    \delta t_c(r)&=\frac{2V_{pd}}{N_r}\sum_{\langle \mathbf{i}_d,\mathbf{j}_d\rangle_r}\sum_{\langle\alpha\beta\rangle_{pd}} w_{\mathbf{i}_d}(\alpha)w_{\mathbf{j}_d}(\alpha)\,w_{\mathbf{i}_d}(\beta)w_{\mathbf{j}_d}(\beta),\\
    V_{\sigma\sigma}(r)&=\delta V_{\uparrow\downarrow}(r)-\delta t_c(r).
    \label{eq:Vpd_corrections}
    \end{aligned}
\end{equation}
In the presence of nearest-neighbor repulsion terms in the three-band model, $t_c\neq V_{\uparrow\downarrow}$ and we acquire an additional $V_{\sigma\sigma}$ interaction term. 

\begin{figure*}[t]
    \centering
    \includegraphics[width=1\textwidth]{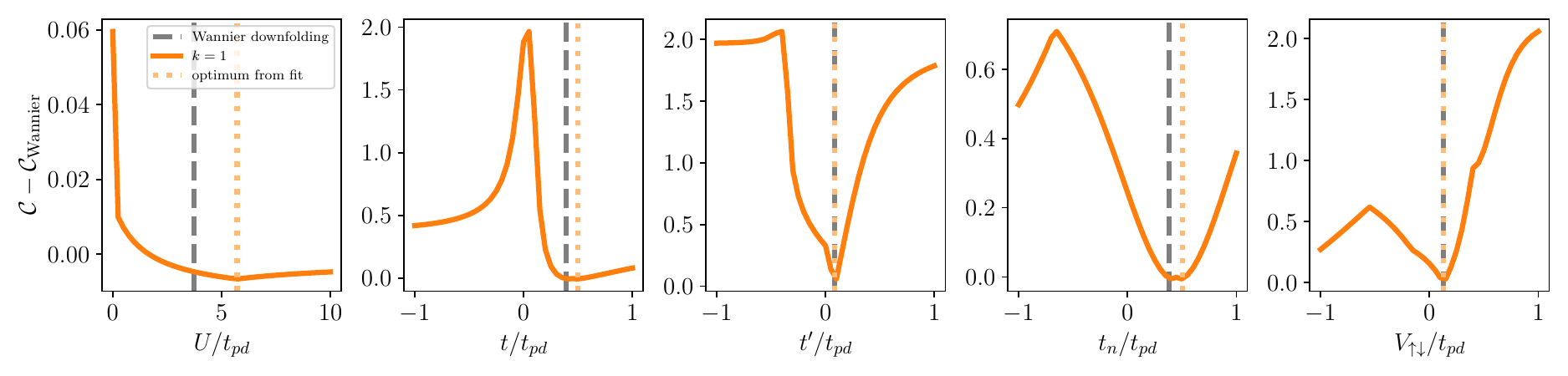} \includegraphics[width=1\textwidth]{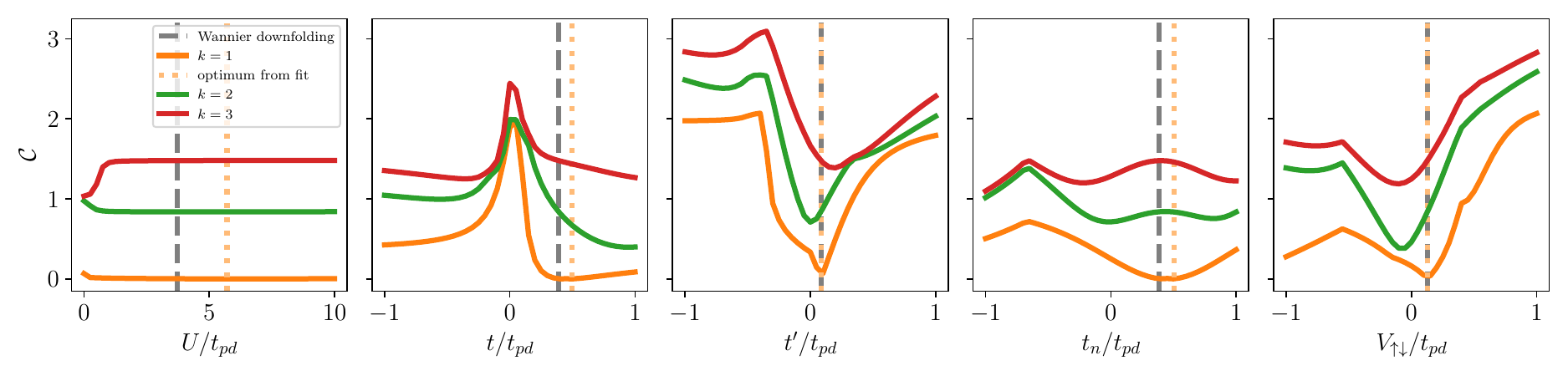}
    \caption{Benchmark of the downfolded single-band model against the low-energy spectrum of the Emery model on a $2\times2$ plaquette with two doped holes. The deviation between the two spectra, $\mathcal{C}-\mathcal{C}_{\mathrm{Wannier}}$, is shown as a function of the individual effective parameters. The Wannier-downfolded values, retaining the five largest effective parameters, are indicated by gray dashed vertical lines. Top: Deviation obtained by comparing only the lowest band ($k=1$). The minima are found close to the Wannier parameters, demonstrating that the downfolded model accurately reproduces the low-energy spectrum. Bottom: Same analysis including multiple bands ($k>1$).
    }
    \label{fig:testdownfolding}
\end{figure*}

\subsection{Comparison of Emery vs. downfolded single-band band structures}

To assess the quality of the reconstruction obtained with the protocol described above, we compare the many-body spectra of the Emery model, $E_n^\mathrm{Emery}(\mathbf{k})$ to the downfolded effective single-band dispersion, $E_n^\mathrm{1b}(\mathbf{k})$ across momenta~$\mathbf{k}$. Hereby $n$ denotes the band index. Specifically, we calculate~\footnote{We find it convenient to additionally rescale $E^{\rm 1b}_n(\mathbf{k})\to s\cdot E^{\rm 1b}_n(\mathbf{k})$,
where the scale $s$ is determined by minimizing $ \vert s\cdot E_n^{\rm 1b}(\mathbf{k}) - E_n^{\rm Emery}\vert ^2$ with fixed (Wannier downfolded) single- and three-band parameters.}
\begin{align}
    \mathcal{C} = \sum_{n=0}^{k-1}\sum_{ \mathbf{k}} \big( E_n^{\rm 1b}(\mathbf{k}) - E_n^{\rm Emery}(\mathbf{k})\big)^2
\end{align}
We consider a plaquette of $2\times 2$ unit cells with two doped holes, allowing us to obtain the spectra via exact diagonalization.

Fig.~\ref{fig:testdownfolding} demonstrates that the effective single-band Fermi-Hubbard model with parameters determined using the Wannier downfolding protocol approximates the low-energy spectrum to a high accuracy when the largest five effective parameters are kept. For the following discussion we denote the deviation of the single- vs. three-band spectra with the respective five Wannier-downfolded parameters as $\mathcal{C}_\mathrm{Wannier}$. The top panel shows the deviation w.r.t. this Wannier benchmark, $\mathcal{C}-\mathcal{C}_\mathrm{Wannier}$, when only the lowest band ($k=1$) is considered. We find that the Wannier parameters (gray dashed vertical lines) coincide to a high accuracy with the minima of $\mathcal{C}$. We do not expect exact agreement since the downfolding is no exact transformation. Furthermore, only the $5$ largest parameters are kept and all effective parameters $P$ with $\vert P\vert <0.1t_{pd}$ have been neglected here. The largest deviation is found for $U/t_{pd}$ (left panel), where $\mathcal{C}-\mathcal{C}_\mathrm{Wannier}$ is very flat. By design, the deviations become larger when more bands ($k>1$) are considered, with the minima deviating from the Wannier downfolding results. We hence conclude that the downfolded model is a good approximation to the lowest band, but becomes less accurate when higher bands are considered.

\section{NQS Benchmarks}
\label{appendix:benchmarks}
In this section, we provide additional benchmarks on the performance of our NQS architecture. 
\subsection{Comparison to DMRG}

\begin{figure*}[t]
    \centering
    \includegraphics[width=0.7\textwidth]{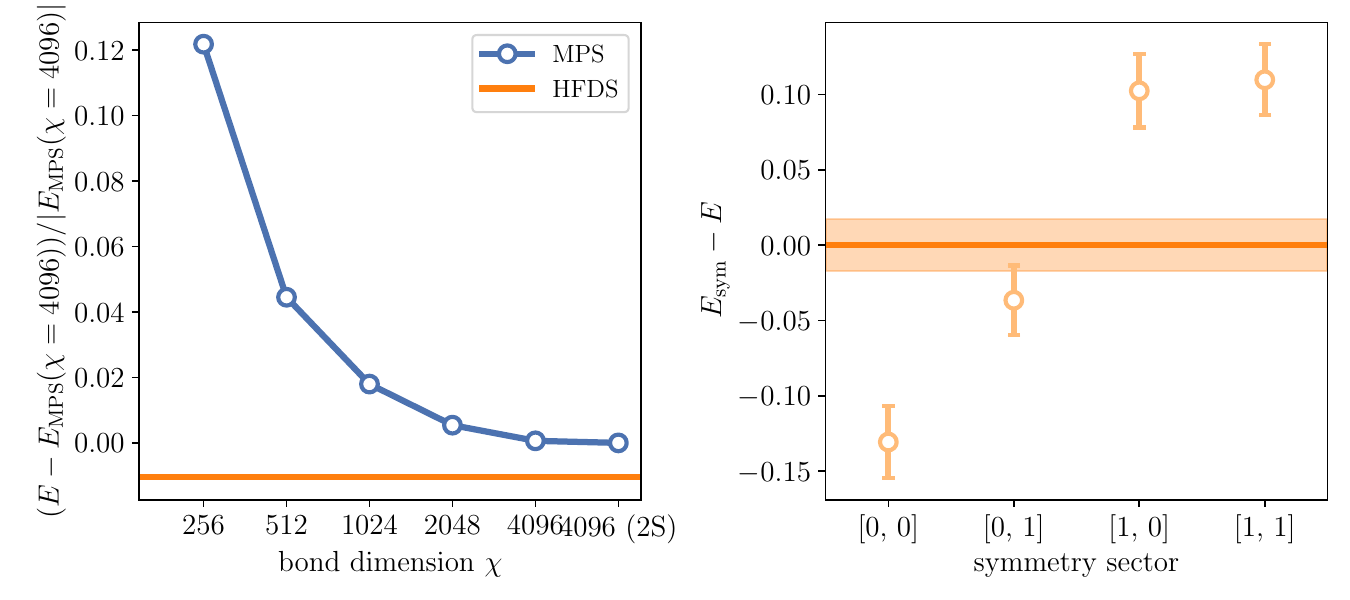}
    \caption{MPS vs HFDS energies for a $10\times 4$ system at $\delta=1/8$ doping. Left: We compare the energies obtained with MPS on a cylinder geometry (blue) using bond dimensions $\chi=256,\dots,4096$ and single-site (two-site) updates for the first five (the last) stages with HFDS on a torus (orange line) with $f=60$ features, $3$ layers, kernel size $3$ and $N_h=10$ hidden fermions. Right: Energies from HFDS without (line) and with translation (by half the system size) symmetry (markers). 
    }
    \label{fig:Energies10x4}
\end{figure*}

Figure~\ref{fig:Energies10x4} compares the variational energies obtained with HFDS to DMRG/MPS results for a $10\times4$ system at $\delta=1/8$ doping. Note that as in the main text different boundary conditions are employed in the two approaches, yielding lower energies for the HFDS due to the additional $4$ bonds in the long direction. We further find that imposing translation symmetry (by half a system size) yields only a small change in the optimized energy. 

\begin{figure*}[t]
    \centering
    \includegraphics[width=0.8\textwidth]{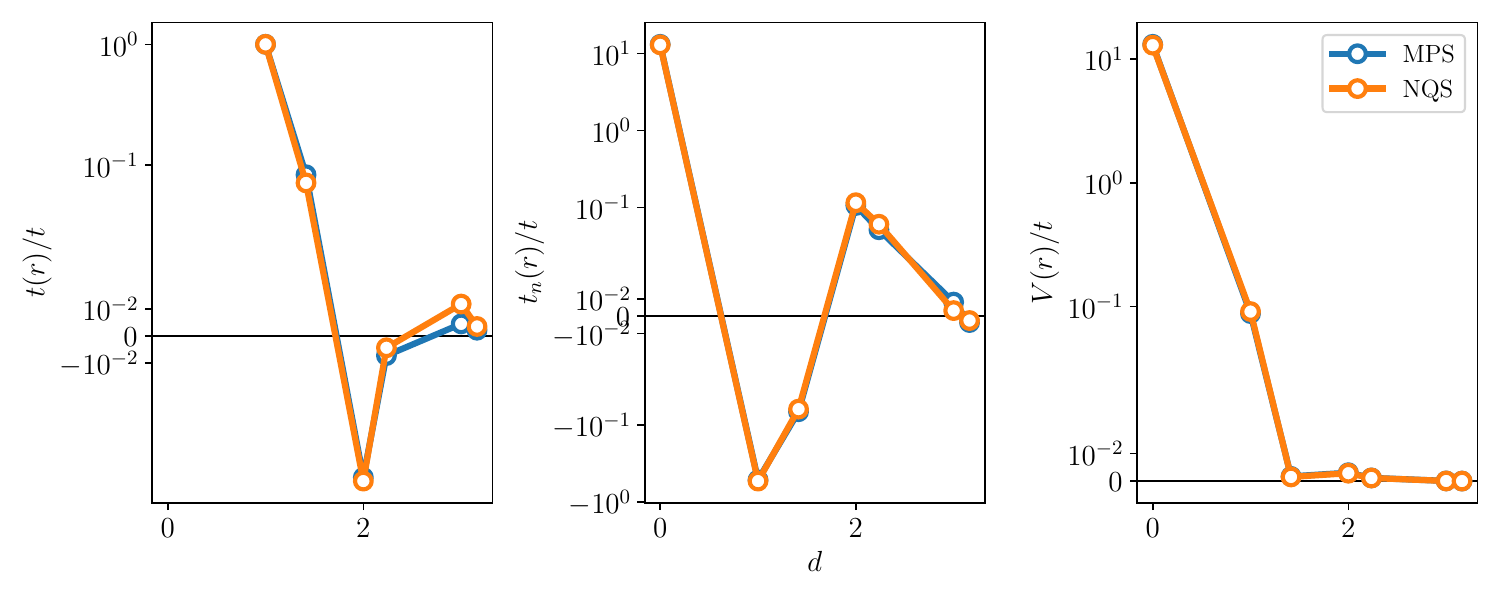}
    \caption{Reconstructed effective single-band parameters obtained with MPS (blue) and HFDS (orange) for a $10\times 4$ system at $\delta=1/8$ doping. Note that we do not expect exact agreement due to the different boundary conditions for MPS and HFDS.
    }
    \label{fig:SinglebandParams10x4}
\end{figure*}

Figure~\ref{fig:SinglebandParams10x4} shows the corresponding reconstructed effective single-band parameters. The effective parameters obtained with HFDS and MPS are in very good agreement, and only small differences visible for $t(r>2)$, where the parameters themselves are very small. Note that small differences are expected due to the different boundaries employed in the two calculations.

\subsection{HFDS Scans}
\begin{figure*}[t]
    \centering
    \includegraphics[width=0.7\textwidth]{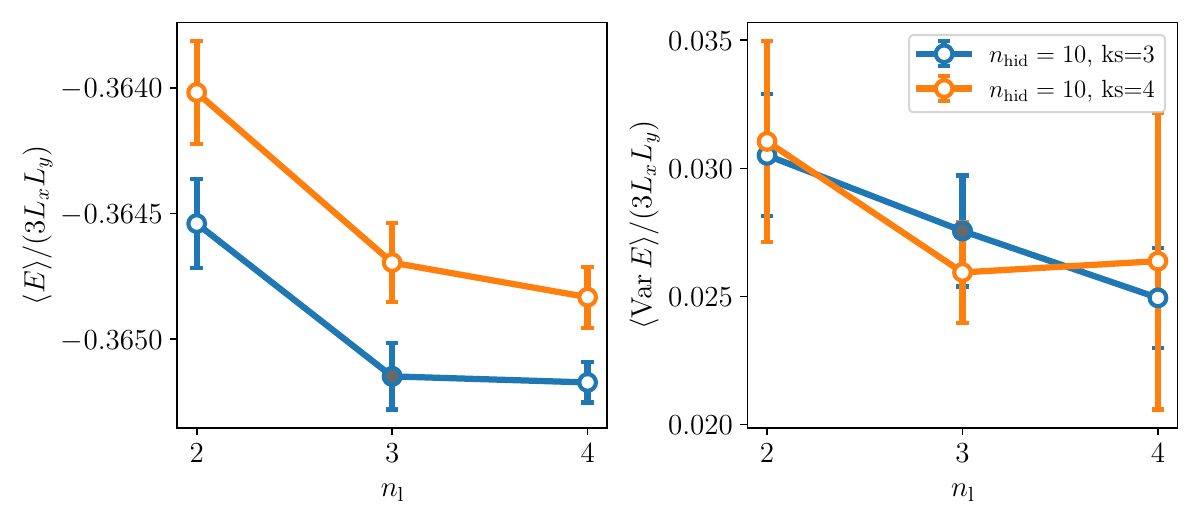}
    \caption{Variational energies and variance of a system with $6\times6$ unit cells, shown for a given number of hidden layers $n_\text{l}$ and two different kernel sizes ks.}
    \label{fig:scan_nlayers}
\end{figure*}

\begin{figure*}[t]
    \centering
    \includegraphics[width=0.7\textwidth]{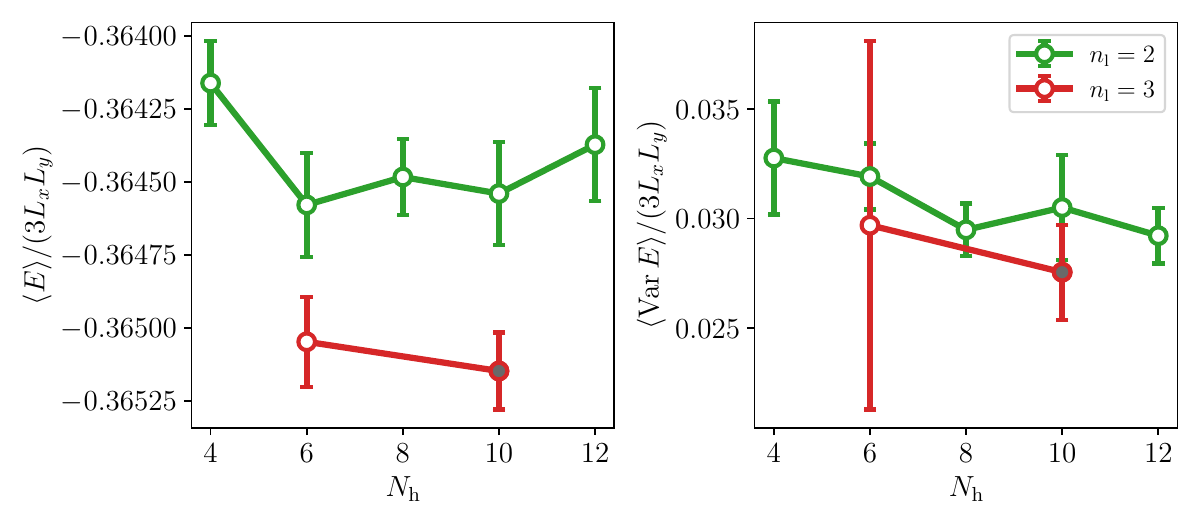}
    \caption{Variational energies and variance of a system with $6\times6$ unit cells, shown for a given number of hidden fermions $N_\text{h}$ and two different $n_\text{l}$.
    }
    \label{fig:scan_nhid}
\end{figure*}

The variational parameters of the HFDS-architecture have been determined via hyperparameter-scans. We present an exemplary scan for a system of $6\times6$ unit cells below, where we compare the variational energies $\langle E\rangle$ and variance $\langle \text{Var}\ E\rangle$ after convergence without imposed symmetries. The three most impactful quantities are determined by the CNN-architecture and the hidden fermion construction:
\begin{itemize} 
    \item ks: Kernel size of the CNN, scan seen in Fig. \ref{fig:scan_nlayers} with $N_\text{h}=10$ fixed. 
    \item $n_\text{l}$: Number of hidden layers, scan seen in Fig. \ref{fig:scan_nlayers} with $N_\text{h}=10$ fixed.
    \item $N_\text{h}$: Number of hidden fermions, scan seen in Fig. \ref{fig:scan_nhid} with $\text{ks}=3$ fixed.
\end{itemize}
The set of variational parameters that yields the lowest energy (dark marker in Figure \ref{fig:scan_nlayers} and \ref{fig:scan_nhid}) has been chosen for our simulations and referenced in section \ref{sec:methods}.\\
We note that recently, constructions employing Pfaffians have been proposed as an alternative to the Slater-determinant-based approaches. We have incorporated Hidden Fermion Pfaffians States (HFPS)~\cite{chen2025neuralnetworkaugmentedpfaffianwavefunctions} in our scans, and found that for a comparable number of variational parameters, constructions with HFDS led to lower variational energies.

\FloatBarrier
\subsection{Comparison to mean-field results}

\begin{figure*}[htp]
    \centering
    \includegraphics[width=0.98\textwidth]{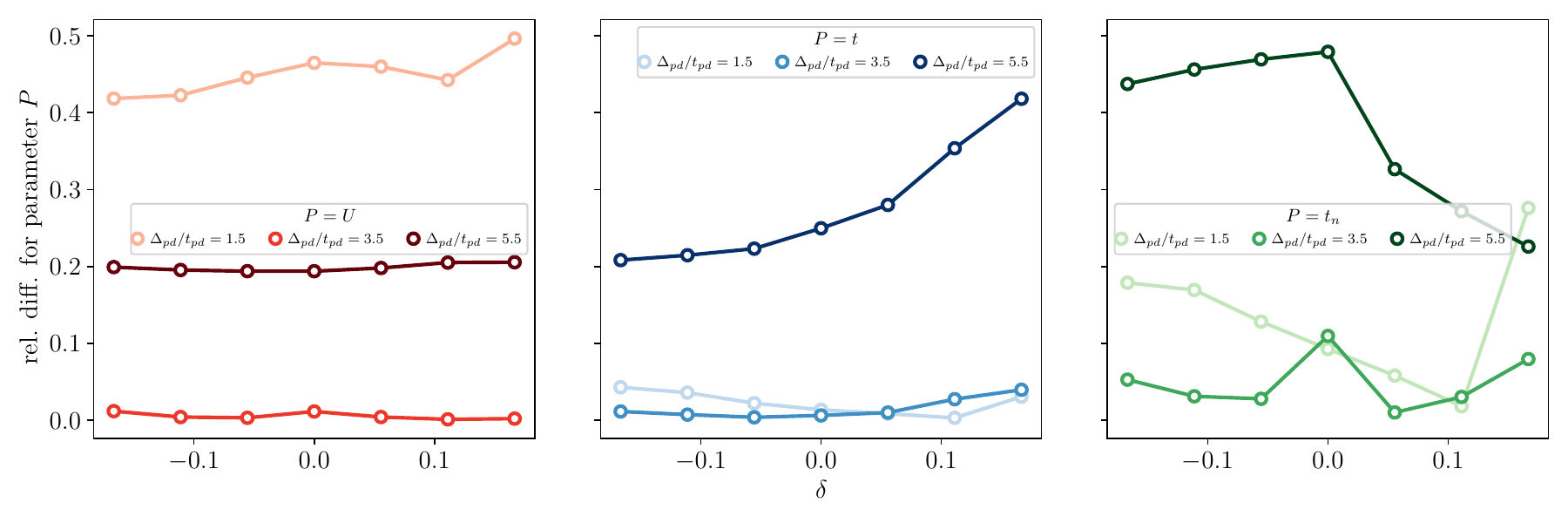}
    \caption{Relative difference between effective parameters obtained from non-interacting mean-field vs. interacting HFDS calculations on $8\times 8$ systems at $\delta=1/8$ doping. For the non-interacting simulation, we employ a $U_\alpha^\mathrm{MF}$ on the mean-field level. 
    }
    \label{fig:MFparams}
\end{figure*}

\begin{figure*}[t]
    \centering
    \includegraphics[width=0.9\textwidth]{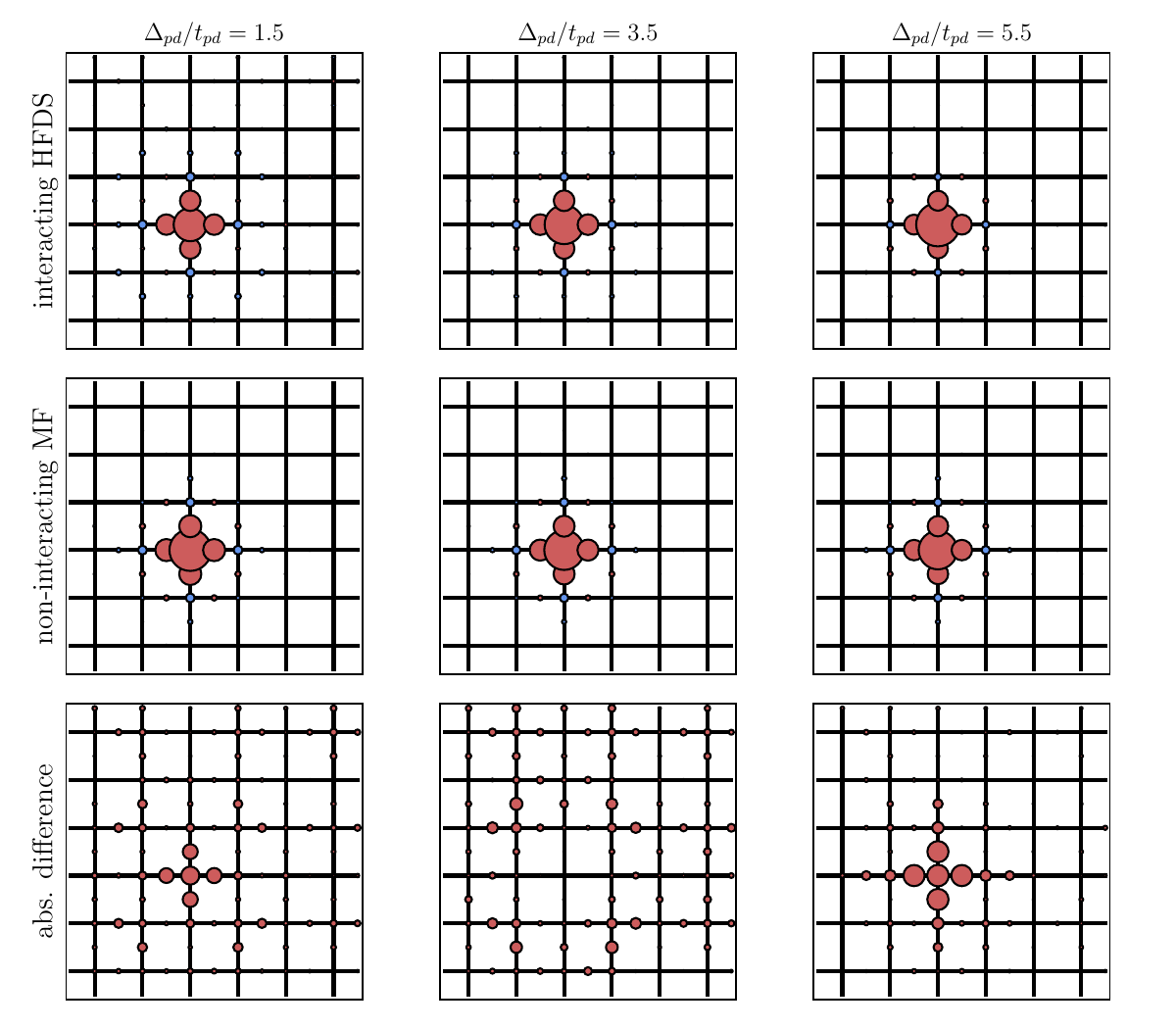}
    \caption{Wannier functions obtained from interacting HFDS (first row) vs. non-interacting mean-field (second row) calculations on $8\times 8$ systems at $\delta=1/8$ doping for different exemplary values of $\Delta_{pd}/t_{pd}$ (columns). The third row displays the difference.
    }
    \label{fig:MFWannier}
\end{figure*}

To assess the role of interactions in the reconstruction of the effective low-energy description, we compare the HFDS results to those obtained from a non-interacting mean-field treatment. Figure~\ref{fig:MFparams} shows the relative differences in the extracted effective single-band parameters. Significant deviations in all three largest effective parameters are present for many of the considered Emery data sets. This is also reflected in the corresponding Wannier functions shown in Fig.~\ref{fig:MFWannier}: although their overall spatial structure is similar, interaction-induced changes in the orbital weight distribution are clearly visible, with the differences depending on $\Delta_{pd}/t_{pd}$.

\subsection{Spin and charge correlations}
\begin{figure*}[t]
    \centering
    \includegraphics[width=0.98\textwidth]{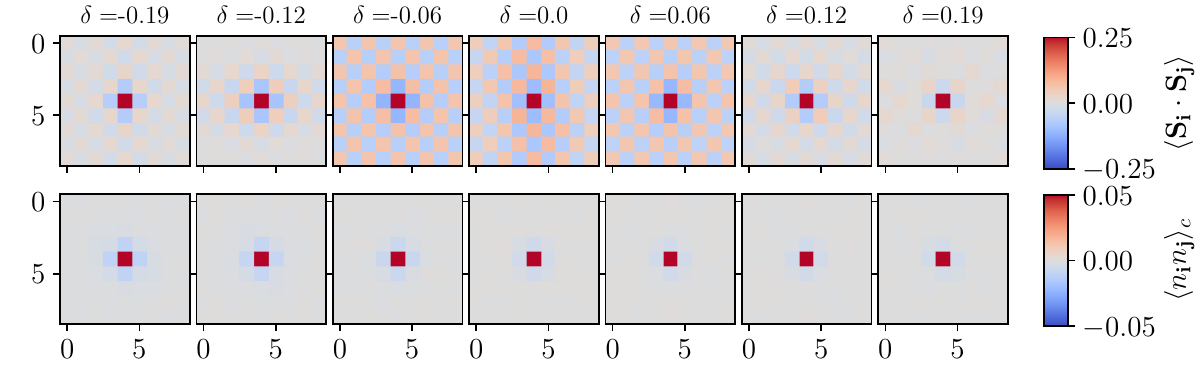}
    \caption{Spin and charge correlation maps for different dopings $\delta$.
    }
    \label{fig:ssnn}
\end{figure*}

In Fig~\ref{fig:ssnn}, we show spin and charge correlations for the $8\times 8$ systems at different fillings and $\Delta_{pd}/t_{pd}=3.5$.

\FloatBarrier
\section{Additional downfolding results}\label{appendix:adddownfolding}
In this section, we provide additional results on the downfolding. We start with the interacting Wannier functions, before turning extended versions of Fig.~5 in the main text with more effective parameters and different dependencies. Lastly, we discuss the effective scaling of the single-band $U$ with the charge-transfer gap.

\subsection{Interacting Wannier orbitals}
\begin{figure*}[htp]
    \centering
    \includegraphics[width=1\textwidth]{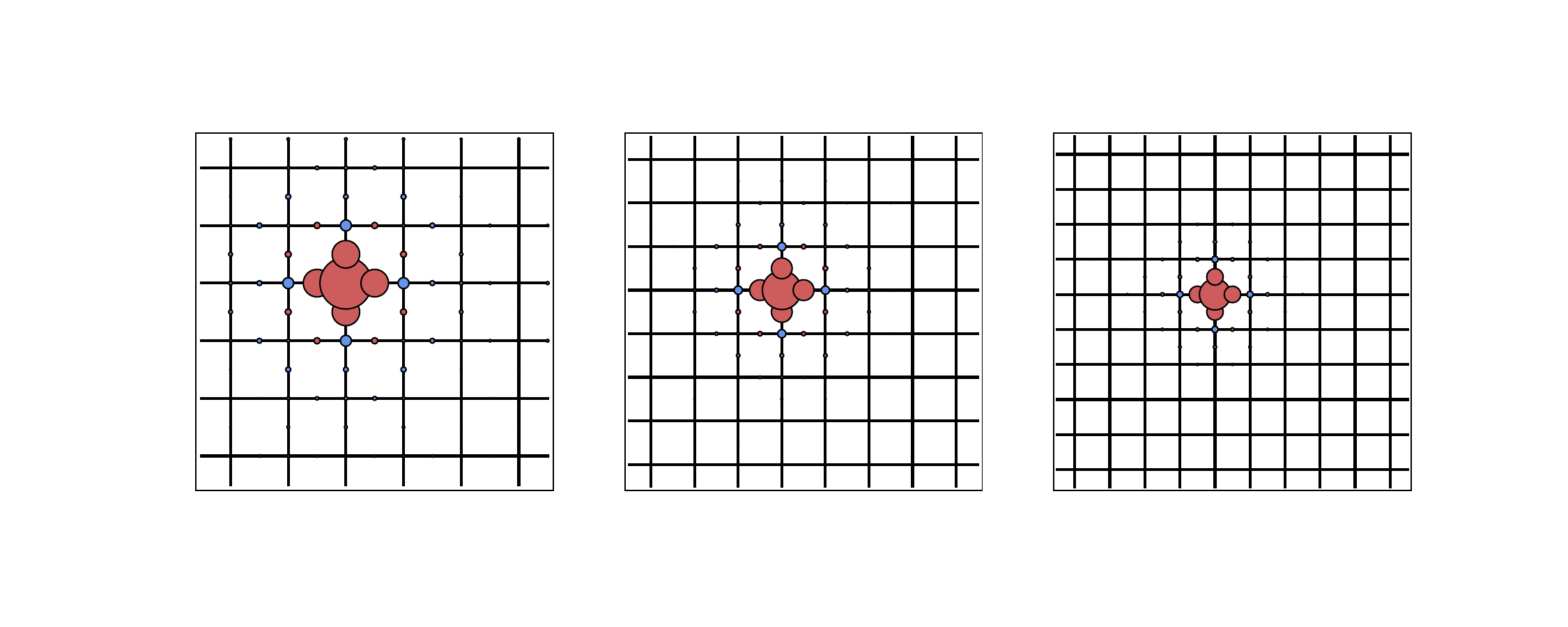}
    \caption{Interacting Wannier functions for $6\times 6$, $8\times 8$ and $10\times 10$ systems at $\delta\approx 1/8$ and $\Delta_{pd}/t_{pd}=3.5$.
    }
    \label{fig:Wannieroverview}
\end{figure*}

Fig.~\ref{fig:Wannieroverview} displays the interacting Wannier functions for different system sizes: $6\times 6$, $8\times 8$ and $10\times 10$, all at $\delta\approx 1/8$ and $\Delta_{pd}/t_{pd}=3.5$.

\begin{figure*}[t]
    \centering
    \includegraphics[width=0.49\textwidth]{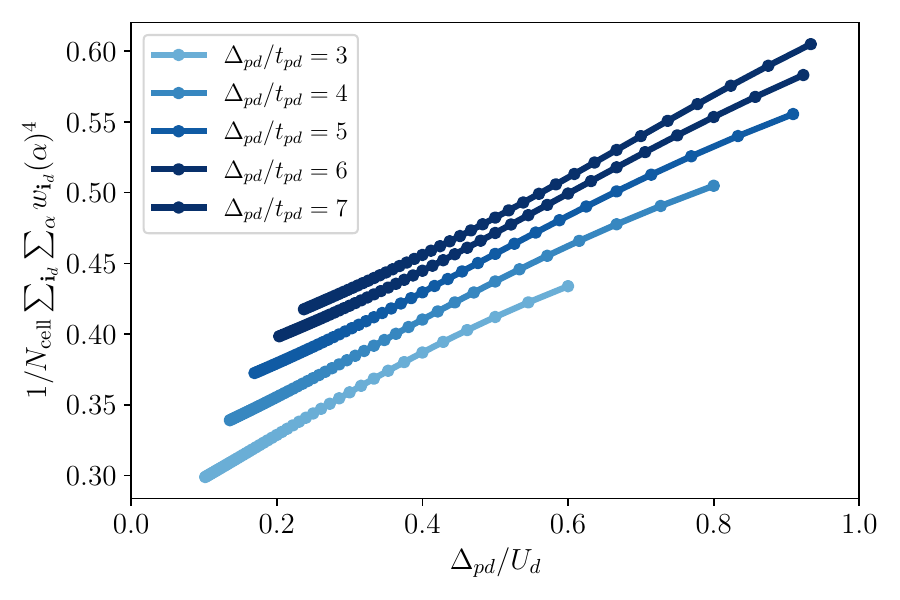}\includegraphics[width=0.49\textwidth]{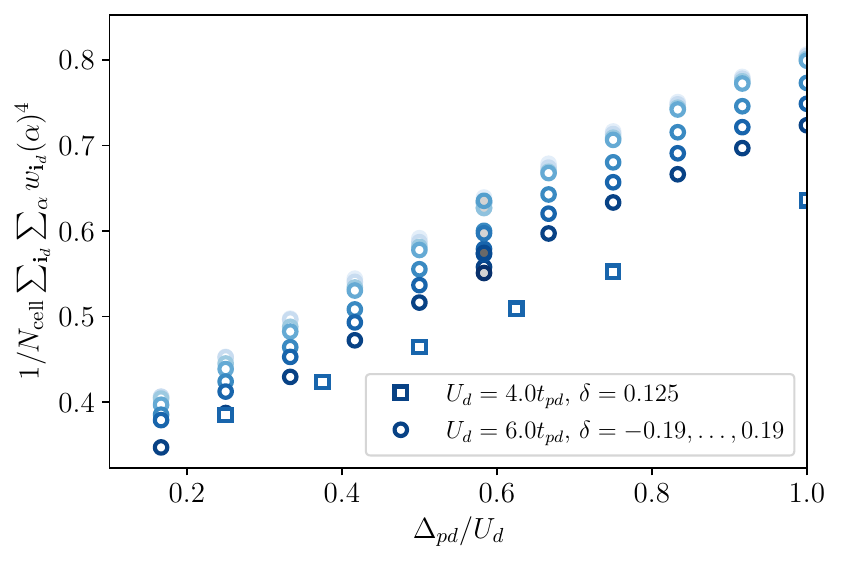}
    \caption{Establishing the dependence $\sum_{\mathbf{i}_d}\sum_\alpha w_{\mathbf{i}_d}(\alpha)^4\propto \Delta_{pd}/U_d$. Left: The Wannier contribution $\sum_{\mathbf{i}_d}\sum_\alpha w_{\mathbf{i}_d}(\alpha)^4$ from exact diagonalization of a $2\times 2$ cluster with two dopants for $U_p=0$, various $U_d$ and $\Delta_{pd}$. A clear linear dependence is visible for $\Delta_{pd}<U_d$. Right: The linear dependence is also visible in our HFDS dataset.
    }
    \label{fig:w^4vsDeltaU}
\end{figure*}
\subsection{The dependence of the effective $U$ on the charge-transfer gap $\Delta_{pd}/t_{pd}$}
As shown in Fig.~\ref{fig:paramsoverview2}, third row, the effective $U/t_{pd}$ scales as $U/t_{pd}\propto \Delta_{pd}/t_{pd}$. This is expected since in the large-$U_d$ (charge-transfer) regime it is energetically more favorable to remove an electron from oxygen than from copper, so the relevant energy cost setting the effective $U$ is controlled by the charge-transfer gap $\Delta_{pd}$ rather than by $U_d$ (see e.g. Refs.~\cite{Hybertsen1992,Zheng2018FromRealMaterialstoModelHamiltonians}). This is not directly apparent from the transformation Eq.~\eqref{eq:trafo}, and the resulting expression for $U$, Eq.~\eqref{eq:effU}, which is reprinted here for clarity:
\begin{equation}
U/t_{pd}=\frac{1}{N_\mathrm{cell}}\sum_{\mathbf{i}_d}\sum_\alpha (U_\alpha/t_{pd})\,  w_{\mathbf{i}_d}(\alpha)^4.
\end{equation}
Instead, the scaling $U/t_{pd}\propto \Delta_{pd}/t_{pd}$ results from the scaling of the respective contributions $\sum_{\mathbf{i}_d}\sum_\alpha w_{\mathbf{i}_d}(\alpha)^4$: Fig.~\ref{fig:w^4vsDeltaU} demonstrates that both for the ground state of a $2\times 2$ cluster obtained with ED as well as our HFDS parameter sets,  $\sum_{\mathbf{i}_d}\sum_\alpha w_{\mathbf{i}_d}(\alpha)^4\propto \Delta_{pd}/U_d$ for $\Delta_{pd}<U_d$.

\subsection{Extended version of Fig.~5}
\begin{figure*}[htp]
    \centering
    \vspace{1cm}
    \includegraphics[width=1\textwidth]{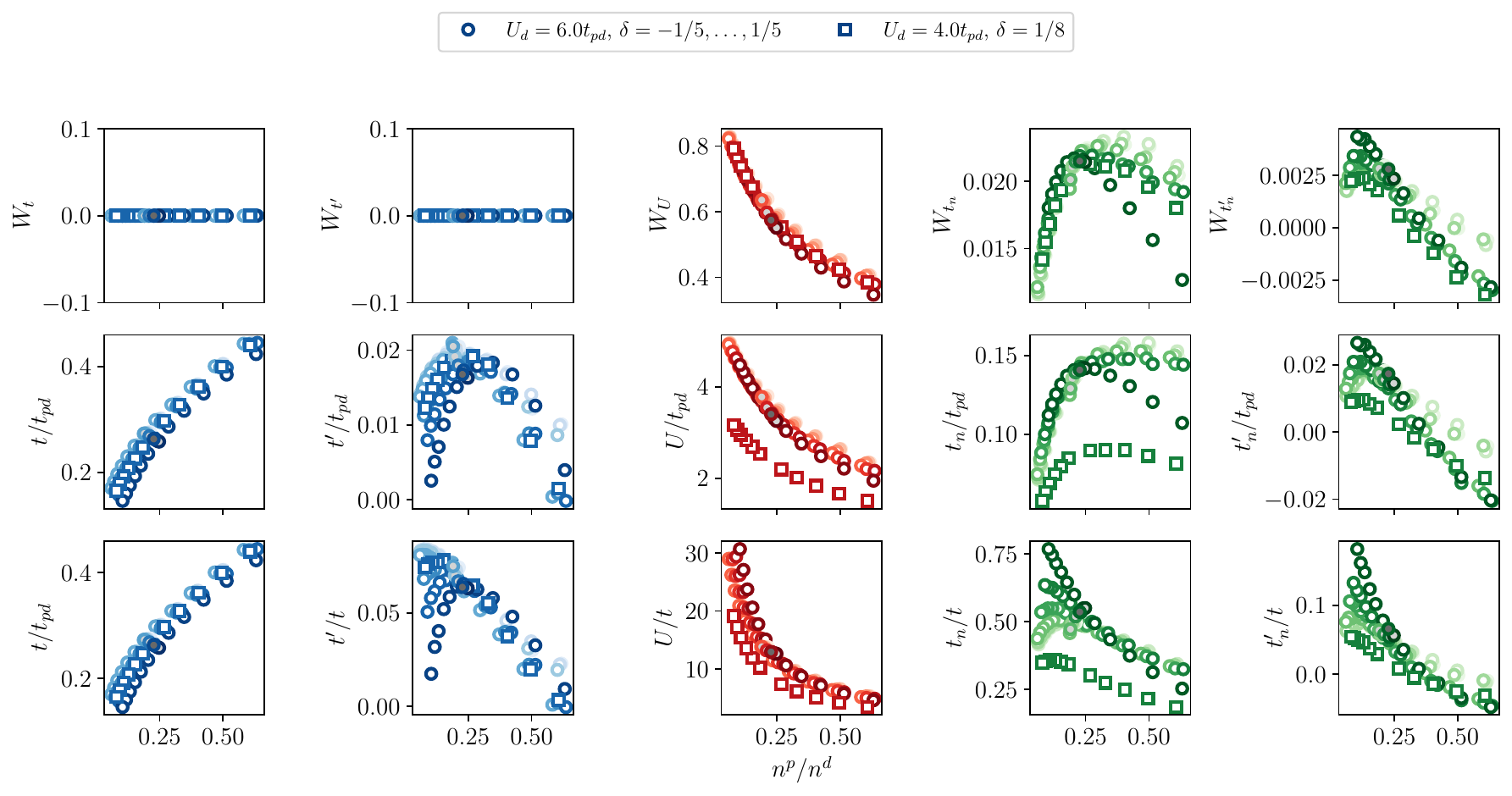}
    \caption{Extended version of Fig.~5b including longer-ranged effective parameters and a middle column with the parameters in units of $t_{pd}$.
    The effective parameters and the Wannier contributions $W$ are plotted as a function of the relative $p$- vs. $d$-site densities, $n^p/n^d$. As in Fig.~5, we plot different data groups: the cuprate (nickelate)-inspired regime with $U_d/t_{pd}=6.0\, (4.0)$, varying the charge-transfer energy $\Delta_{pd}/t_{pd}$ and doping $\delta$, with the latter denoted by light to dark colors. System sizes $6\times 6$ ($8\times 8$, $10\times 10$) are indicated by white (light gray, dark gray) marker filling.}
    \label{fig:fig5ext}
\end{figure*}

Fig.~\ref{fig:fig5ext} is an extended version of Fig.~5b in the main text, including longer-ranged effective parameters $t^\prime_{(n)}$ and a middle column with the parameters in units of $t_{pd}$.

\subsection{Parameter scans w.r.t. the charge-transfer gap $\Delta_{pd}/t_{pd}$}

\begin{figure*}[htp]
    \centering
    \includegraphics[width=0.92\textwidth]{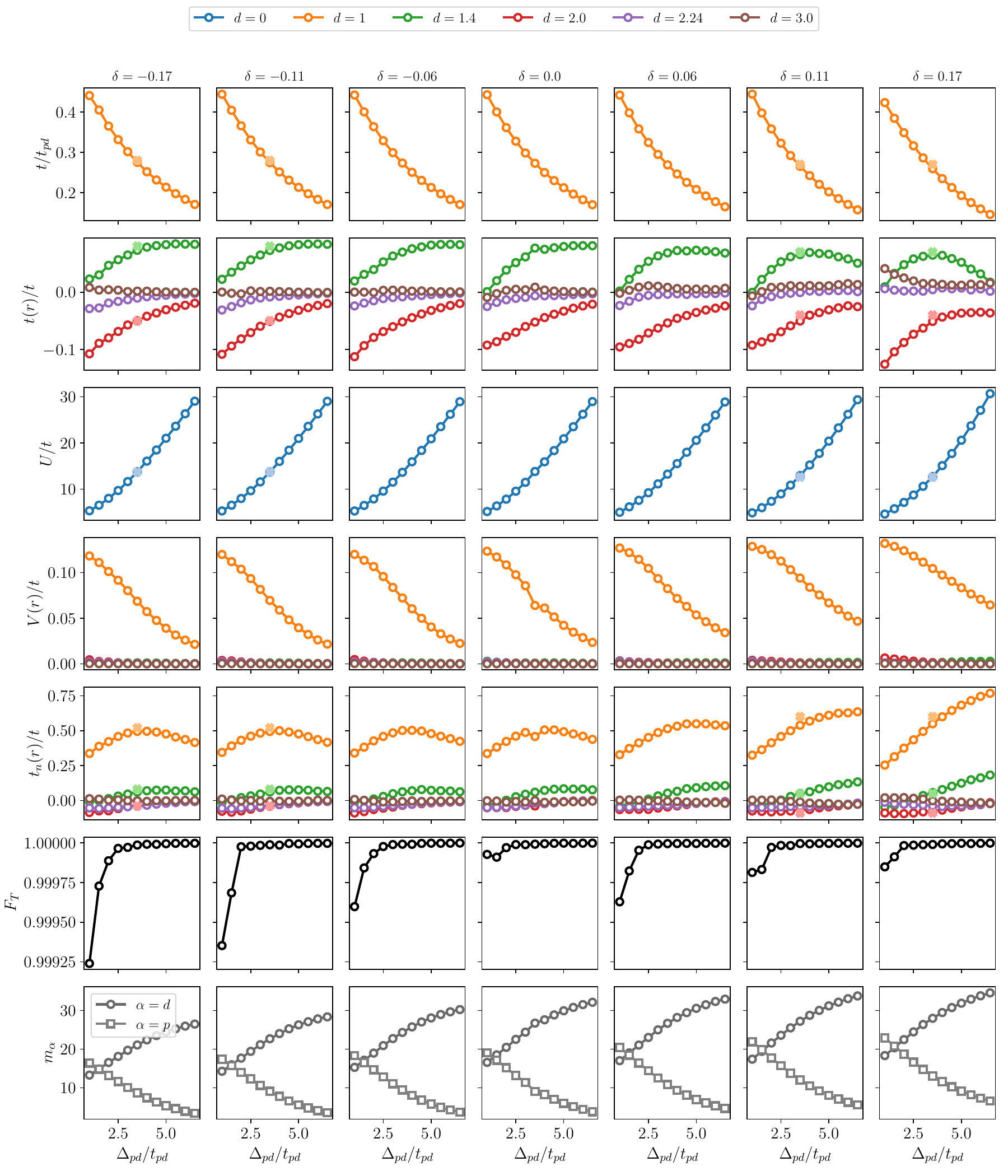}
    \caption{Effective parameters $t(r)$, $U$, $t_n(r)$ and $V(r)$ in units of the effective $t$ for $6\times 6$ systems at different dopings (columns). We furthermore show the overlap with translated Wannier functions, $F_T=1/(N_TN_\mathrm{cell})\sum_{\mathbf{i,T}}\langle w_\mathbf{i}\vert w_\mathbf{i+T}\rangle^2$ and the $\alpha$-mass defined in Eq.~\eqref{eq:mass}. 
    }
    \label{fig:paramsoverview}
\end{figure*}

\begin{figure*}[htp]
    \centering
    \includegraphics[width=0.98\textwidth]{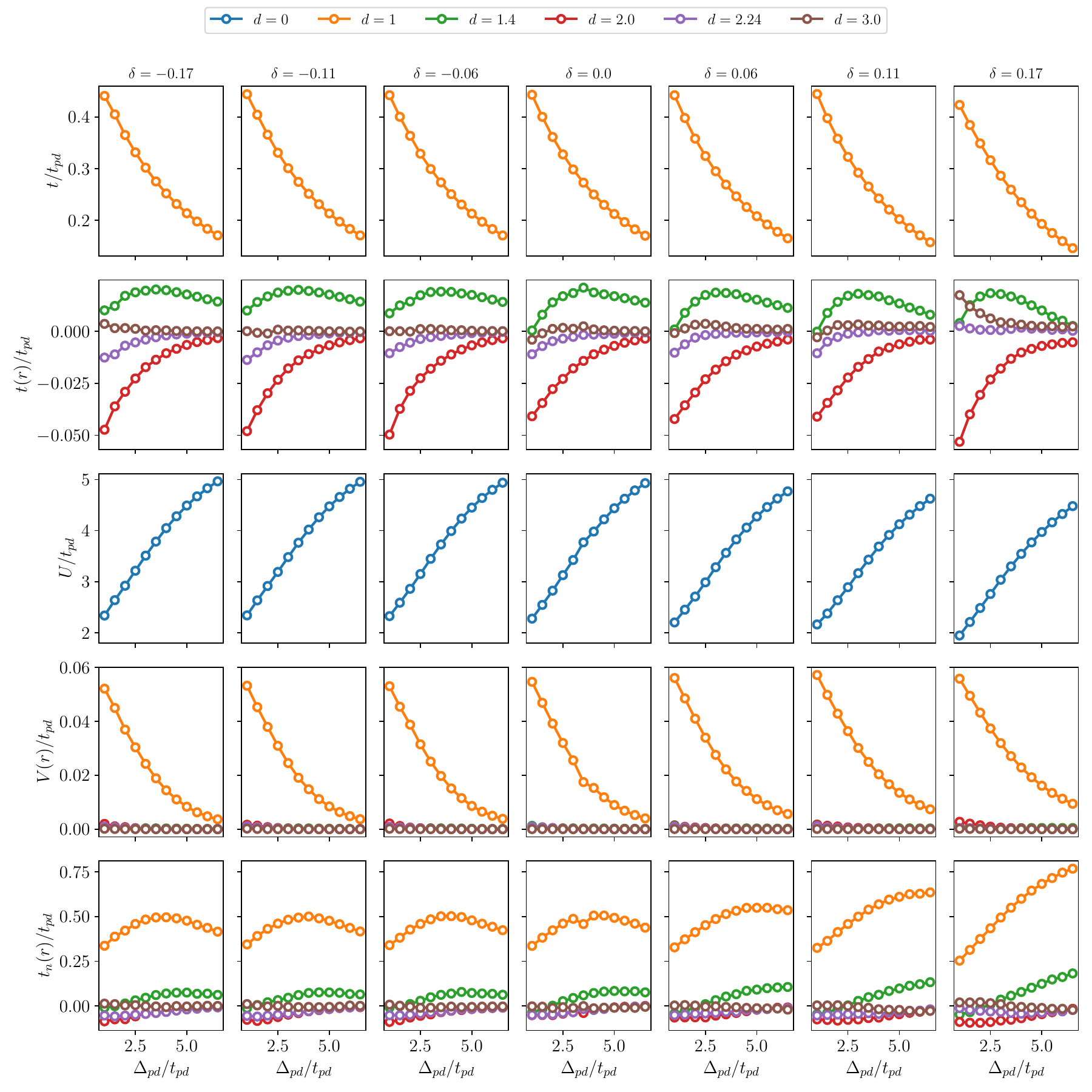}
    \caption{Effective parameters $t(r)$, $U$, $t_n(r)$ and $V(r)$ in units of the copper-oxygen hopping strength $t_{pd}$ for $6\times 6$ systems at different dopings (columns).
    }
    \label{fig:paramsoverview2}
\end{figure*}

\end{document}